\documentclass[times,twocolumn]{aastex63}

\defcitealias{Ghosh_Lamb1979}{GL79}
\defcitealias{2018ApJ...863....9W}{WMJ18}

\received{} 
\revised{} 
\accepted{} 

\shorttitle{Swift J0243.6+6124 with MAXI GSC}
\shortauthors{Sugizaki et al.}

\begin{document} 

\title{
X-ray emission evolution of
the Galactic ultra-luminous X-ray pulsar 
Swift J0243.6+6124 
during the 2017--2018 outburst
observed by the MAXI GSC
}

\email{mutsumi@nao.cas.cn}

\author[0000-0002-1190-0720]{Mutsumi Sugizaki} 
\affiliation{Department of Physics, Tokyo Institute of Technology, 
2-12-1 Ookayama, Meguro-ku, Tokyo 152-8551, Japan}
\affiliation{National Astronomical Observatories, Chinese Academy of Sciences, 
20A Datun Rd, Beijing 100012, China}

\author{Motoki Oeda}
\affiliation{Department of Physics, Tokyo Institute of Technology, 
  2-12-1 Ookayama, Meguro-ku, Tokyo 152-8551, Japan}

\author[0000-0001-9656-0261]{Nobuyuki Kawai}
\affiliation{Department of Physics, Tokyo Institute of Technology, 
  2-12-1 Ookayama, Meguro-ku, Tokyo 152-8551, Japan}

\author[0000-0002-6337-7943]{Tatehiro Mihara}
\affiliation{High Energy Astrophysics Laboratoy, RIKEN, 2-1 Hirosawa, Wako, Saitama 351-0198, Japan}

\author{Kazuo Makishima}
\affiliation{High Energy Astrophysics Laboratoy, RIKEN, 2-1 Hirosawa, Wako, Saitama 351-0198, Japan}
\affiliation{Kavli IPMU, The University of Tokyo, 5-1-5 Kashiwanoha, Kashiwa, Chiba 277-8583, Japan}

\author{Motoki Nakajima}
\affiliation{School of Dentistry at Matsudo, Nihon University, 
  2-870-1 Sakaecho-nishi, Matsudo, Chiba 101-8308, Japan}


\begin{abstract}
This paper reports on the X-ray emission evolution of the ultra-luminous Galactic 
X-ray pulsar, Swift J0243.6+6124, during the giant outburst  from 2017 October to 2018 January 
as observed by the MAXI GSC all-sky survey. 
The 2--30 keV light curve and the energy spectra confirm 
that the source luminosity $L_\mathrm{X}$ assuming an isotropic emission 
reached $2.5\times 10^{39}$ erg s$^{-1}$, 
10 times higher than the Eddington limit for a $1.4 M_\odot$ neutron star. 
When the source was luminous with $L_\mathrm{X}\gtrsim 0.9\times 10^{38}$ erg s$^{-1}$,
it exhibited generally  a negative correlation on a hardness-intensity diagram.
However, two hardness ratios, 
a soft color ($=$ 4--10 keV / 2--4 keV) and a hard color ($=$ 10--20 keV / 4--10 keV),
showed somewhat different behavior
across a characteristic luminosity of $L_\mathrm{c}\simeq 5\times 10^{38}$ erg s$^{-1}$.
The soft color changed more than the hard color when $L_\mathrm{X} < L_\mathrm{c}$,
whereas the opposite was observed above $L_\mathrm{c}$.
The spectral change  above $L_\mathrm{c}$ was represented by a broad enhanced
feature at $\sim 6$ keV on top of the canonical cutoff power-law continuum.
The pulse profiles, derived daily, made a transition from a single-peak 
to a double-peak one as the source brightened across $L_\mathrm{c}$.
These spectral and pulse-shape properties can be interpreted 
by a scenario that the accretion columns on the neutron star surface, 
producing the Comptonized X-ray emission,
gradually became taller as  $L_\mathrm{X}$ increased.
The broad 6 keV enhancement 
could be a result of cyclotron-resonance absorption at $\sim 10$ keV,
corresponding to a surface magnetic field $B_\mathrm{s}\simeq 1.1\times 10^{12}$ G.
The spin-frequency derivatives calculated with the Fermi GBM data 
showed a smooth positive correlation with $L_\mathrm{X}$ up to the outburst peak,
and its linear coefficient is comparable to those of typical Be binary pulsars 
whose $B_\mathrm{s}$ are $(1-8)\times 10^{12}$ G. 
These results suggest that $B_\mathrm{s}$ of Swift J0243.6$+$6124 
is a few times $10^{12}$ G.
\end{abstract}



\section{Introduction}

Swift J0243.6+6124 (hereafter Swift J0243.6)
is a Be X-ray binary pulsar (XBP) discovered on 2017
October 3.
It was first identified as a new X-ray object by
the Swift BAT (Burst Alert Telescope) transient survey \citep{2017GCN21960}.
The MAXI \citep[Monitor of All-sky X-ray Image;][]{Matsuoka_pasj2009}
GSC \citep[Gas Slit Camera;][]{Mihara_pasj2011} 
all-sky monitor also recognized the emergent X-ray activity
almost simultaneously, but could not resolve the source from the nearby
X-ray bianry, LS I $+61$ 303
\citep{2017ATel10803....1S,2017ATel10813....1S}.
The follow-up observations by the Swift XRT (X-ray Telescope) clarified that it is a new
X-ray pulsar with a 9.86 s coherent pulsation
\citep{{2017ATel10809....1K}}.
A timing analysis of Fermi GBM (Gamma-ray Burst Monitor) data
confirmed the periodicity \citep{2017ATel10812....1J},
and also revealed period modulation
due to the binary orbital motion, represented 
by an orbital period of $\sim$ 27 d and an eccentricity of $\sim$ 0.1
\citep{2017ATel10907....1G,2018A&A...613A..19D}.
From optical spectroscopic observations,
the binary companion was identified as a 
Be star \citep{2017ATel10822....1K}.

The long-term X-ray activity of Swift J0243.6 has been continuously
monitored by all-sky X-ray instruments in orbit, ie. the MAXI/GSC,
Swift/BAT, and Fermi/GBM 
\citep[e.g.][]{2018ATel11280....1J, 2018ATel11517....1R}.
The first outburst continued for about 150 d, 
longer than the 27-d orbital period.
The X-ray intensity reached $\sim 5$ Crab at the peak,  
which is comparable to that of the brightest X-ray sources in the sky.
The combined analysis of NICER (Neutron Star Interior Composition Explorer)
and Fermi/GBM data revealed
luminosity-dependent changes both in the hardness ratio and the pulse profile
\citep[][hereafter \citetalias{2018ApJ...863....9W}]{2018ApJ...863....9W}.
%
The X-ray spectrum was also observed repeatedly by pointing X-ray
telescopes including the Swift/XRT, NuSTAR, NICER, and insight-HXMT
\citep[e.g.][]{2019ApJ...873...19T,2019ApJ...879...61Z,2018MNRAS.474.4432J,
2019ApJ...885...18J,
2020MNRAS.491.1857D}.
%
The spectrum was roughly represented by a cutoff power-law continuum and 
an iron-K emission line, 
which agree with those of the typical XBPs \citep{Makishima1999, 2002ApJ...580..394C}.
However, as the source brightened, 
the spectrum began to exhibit a broad enhancement at around 6 keV.
The feature looks like an additional iron-K line with a large width $\sigma\gtrsim1$ keV
\citep{2019ApJ...873...19T,2019ApJ...885...18J}.
Any cyclotron resonance feature due to
the magnetic field on the neutron star surface
has not yet been detected.
Because 
the source intensity became so high, 
the data from the instruments with X-ray mirrors 
were significantly affected by the event pile-up effect 
\citep[][\citetalias{2018ApJ...863....9W}]{2018MNRAS.479L.134T}.

The source distance was first estimated as $D=2.5$ kpc 
from the optical observations of the Be-star companion
\citep{2017ATel10968....1B}.
\citet{2018A&A...613A..19D} derive another estimate, $\sim 5$ kpc,
by applying theoretical accretion-torque models
to the observed 
relation between the X-ray flux and spin-period change.
Lately, in the GAIA DR2 (Data Release 2) based on the purely geometrical method
\citep{2016A&A...595A...1G, 2018A&A...616A...1G},
it has been determined to be 6.8 kpc with a 1-$\sigma$ range of 5.7-8.4 kpc 
\citep{2018AJ....156...58B}. 
This implies that the X-ray luminosity reached $2\times 10^{39}$ erg s$^{-1}$
\citep[][\citetalias{2018ApJ...863....9W}]{2018MNRAS.479L.134T},
10 times higher than the Eddington limit 
for a typical $1.4 M_\odot$ neutron star,
where $M_\odot$ is the Solar mass.
Therefore, the object is 
categorized into an ultra-luminous X-ray pulsar 
\citep[ULXP,][]{2014Natur.514..202B}

Ultra-luminous X-ray sources (ULXs) are defined by the extraordinary high X-ray
luminosities, $\gtrsim 10^{39}$erg s$^{-1}$,
exceeding the Eddington limit 
of typical stellar-mass ($\sim 5M_\odot$) black holes
\citep[e.g.][]{2000ApJ...535..632M,2017ARA&A..55..303K}.
So far, about hundreds of ULXs have been discovered in external galaxies,
although the origin of their extreme luminosity has not yet been understood.
Recently, a few of them were identified as X-ray pulsars, or ULXPs, 
from their coherent X-ray pulsations
\citep{2014Natur.514..202B, 2016ApJ...831L..14F, 2017Sci...355..817I, 2018MNRAS.476L..45C}.
Thus, Swift J0243.6 is a promising candidate for a ULXP, hence a ULX,
that has been found in our Galaxy for the first time.
It provides us a valuable opportunity
to investigate the nature of ULXs.
%
In fact, the X-ray absorption lines detected by the Chandra 
High-Energy Transmission Grating Spectrometer (HETGS) 
from this source
can be explained by a scenario of an ultrafast outflow,
like in the case of other luminous X-ray binaries
\citep{2019MNRAS.487.4355V}.
%
The object is also unique in its significant radio emission,
which is considered as the first evidence of
relativistic jets launched by a slow-rotating, highly-magnetized X-ray pulsar
\citep{2018Natur.562..233V,2019MNRAS.483.4628V},

Since 2009 August, the MAXI GSC on the International Space Station (ISS) 
has been scanning almost the whole sky every 92-minute orbital cycle 
in the 2--30 keV band.
The data have enabled us to study the X-ray evolution of Swift J0243.6
throughout the outburst.
From each transit of the source, lasting 40 s every 92 minutes, 
the GSC provides us with a list of 2--30 keV photons 
with a moderate energy resolution ($\lesssim15$\% at 6 keV)
and a good time resolution (50 $\mu$s),
and the data are free from the event pile-up problem.

The present paper describes the GSC observation and the data analysis of Swift J0243.6
during the giant outburst from 2017 October to 2018 January.
In particular, we focus on the spectral and pulse-profile evolution
around the outburst peak when the luminosity exceeded the Eddington limit.
We also analyze the relation between the X-ray luminosity 
and the pulse-period change
by incorporating the Fermi/GBM pulsar data,
and then discuss possible origins of the unusually high X-ray luminosity
by comparing
with more ordinary XBPs.
In the following analysis,
we employ the orbital parameters as listed in table \ref{tab:orbpar}
that were
first obtained by \citet{2018ATel11280....1J}
and then refined by the Fermi/GBM pulsar analysis\footnote{\url{https://gammaray.msfc.nasa.gov/gbm/science/pulsars.html}},
and $D=7$ kpc from the GAIA DR2 \citep{2018AJ....156...58B}.

\begin{deluxetable}{ll}
\tablecaption{Swift J0243.6 orbital parameters.\label{tab:orbpar}}
\tablehead{
\colhead{Parameter name} &
\colhead{Value} 
}
\startdata
Orbital period $P_\mathrm{orb}$ 	& 27.70 d \\
Projected semi-major axis $a_\mathrm{x}\sin i$ &  115.53 lt-s \\
Eccentricity $e$   		& 0.103 \\
Epoch for mean longitude $90^\circ$ $T_{90}$ & 58115.597 (MJD) \\
Orbital longitude $\omega$ at $T_\mathrm{per}$  & $115.53^\circ$\\
\enddata
\end{deluxetable}

\section{Observation and data reduction}
\label{sec:obs}

We utilized the standard GSC event data 
reduced from the data transferred via
the medium-bit-rate downlink path
in the 64-bit mode.
Because these data are not processed
with any data reduction or event filtering, 
the full 2--30 keV energy range and the 50-$\mu$s
time precision are available
\citep{Mihara_pasj2011}.
We employed the standard analysis tools developed for the instrument calibration
\citep{Sugizaki_pasj2011}.
For each scan transit, 
the source event data were collected from
a rectangular region of $3\fdg 0$  in the scan direction
and $4\fdg 0$ in the anode-wire direction,
with its centroid located at the position of Swift J0243.6.
The backgrounds included in the region
were estimated from the
events in the same detector area,
taken before / after the scan transits.

During the in-orbit operation for over 8 years since 2009,
some of the GSC gas counters out of the 12 units
had already degraded by 2017.
Specifically, 
three units  (GSC\_3, GSC\_6, and GSC\_9)
are operated with their effective area halved.
Furthermore, 
their background rates are 5--10 times higher 
because their anti-coincidence background rejections are disabled.
Another unit, GSC\_1, has been in
a test operation with 
an exceptionally reduced high voltage 
(1500 V versus the normal value of 1650 V).
In addition, GSC\_0 has been suffering gas leak since 2013 June.
Although these gas counters have large response uncertainties, 
the 50 $\mu$s event timing is retained.
We thus use data of these degraded 5 units 
only for the light curve and pulsar timing analysis,
and exclude them from the spectral analysis.

\section{Analysis and results}

\subsection{Light curves and hardness ratios}
\label{sec:analc}

\begin{figure*}
\centering
\includegraphics[width=17cm]{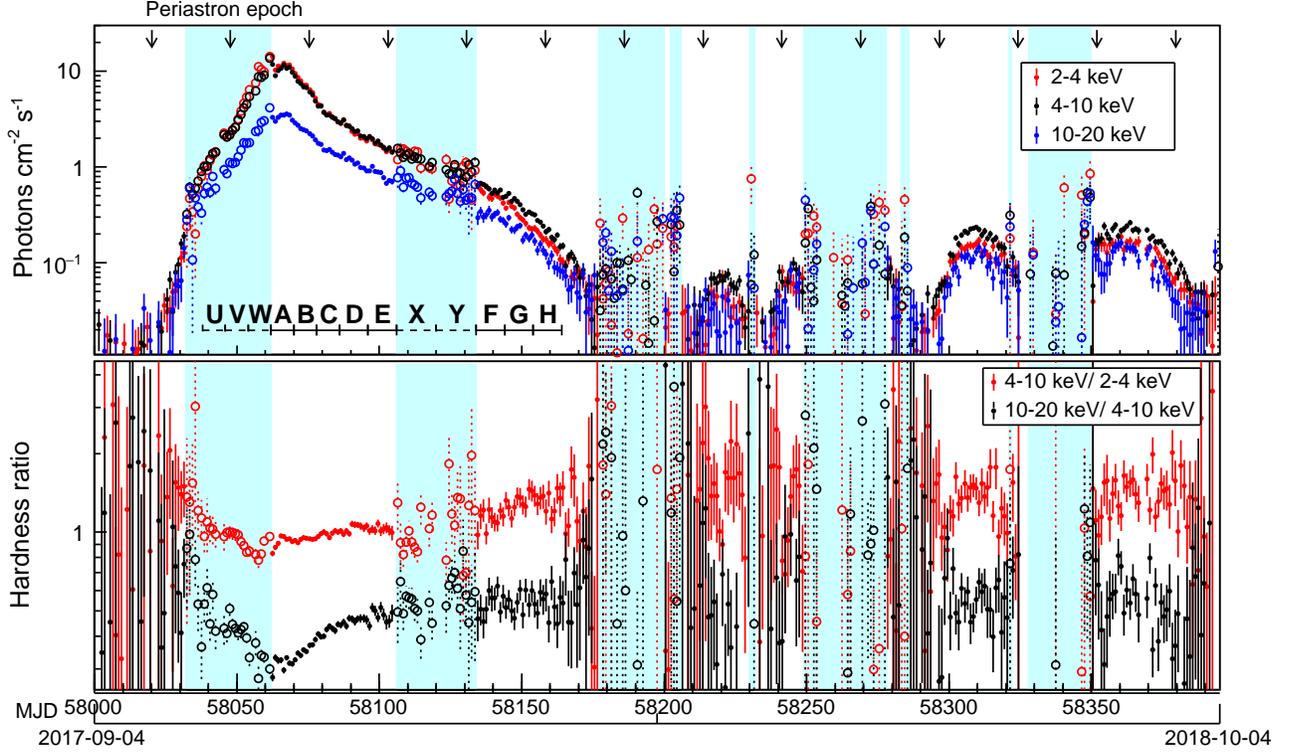} 
\caption{
(Top) 
Background-subtracted X-ray light curves 
of Swift J0243.6 
by the MAXI GSC
in the 2--4 keV, 4--10 keV, and 10--20 keV bands. 
(Bottom) Time variations of two hardness ratios, 
4--10 keV / 2-4 keV 
and 10--20 keV / 4--10 keV. 
Arrows in the top panel represent the epochs of periastron
every 27.6 d orbital period.
The 13 GSC data intervals, U, V, W, A, B, C, D, E, X, Y, F, G, and H, defined in 
table \ref{tab:outint}, are also presented.
In both panels, cyan strips represent the periods covered only by 
the degraded GSC units. 
Data marked with filled ($\bullet$) and open ($\circ$) circles 
are taken by the normal and degrade units,
respectively.
\label{fig:gsclc}
}
\end{figure*}

Figure \ref{fig:gsclc} shows 
the background-subtracted  X-ray light curves of Swift J0243.6
from 2017 September to 2018 October,
obtained by the GSC in the 2--4 keV, 4--10 keV and 10--20 keV bands
in an 1-d time bin.
Also plotted are the time variations of the Soft Color (hereafter SC),
i.e. the 4--10 keV to 2--4 keV intensity ratio,
and the Hard Color (HC), i.e. the 10--20 keV to the 4--10 keV intensity ratio.
These ratios have been calculated after the background subtraction.
To visualize the quality of the degraded units 
(GSC\_0, GSC\_3, and GSC\_6), we plot their data with different symbols.
The statistical errors of these units
are larger typically by a factor of 5--10
than those of the normal units.
%
%

Figure \ref{fig:gsclc} reveals that the present X-ray activity
started at around MJD 58025 (2017 September 29) and continued for
over 1.5 years.  The first outburst developed into the largest one
with the highest peak intensity 
(25 photons cm$^{-2}$ s$^{-1}$ $\simeq 7$ Crab in 2--20 keV) 
and the longest duration ($\sim 150$ d).
After this, several outbursts with lower peaks 
($\lesssim 1$ photons cm$^{-2}$ s$^{-1}$) and 
shorter durations ($\lesssim 40$ d) followed.
Their recurrence cycles do not synchronize with the 27.3-d orbital period.
This means that they are classified into the giant (type-II)
outbursts of Be XBPs \citep[e.g.][]{Reig2011}.

\begin{figure}
\centering
\vspace{5mm}
\includegraphics[width=8.4cm]{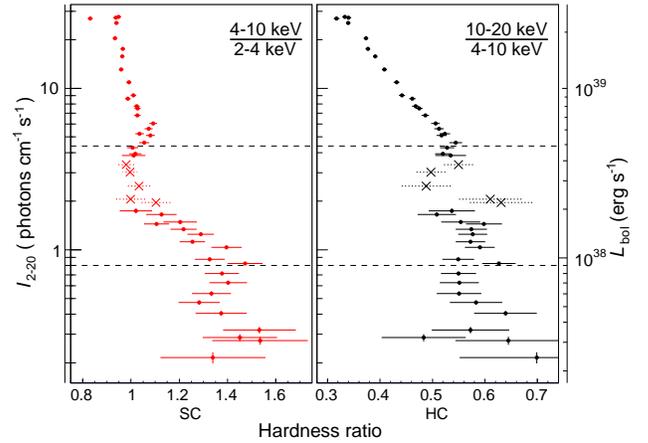} 
\caption{
Hardness-intensity diagrams for 
the SC (left panel) 
and the HC (right panel).
Data marked with crosses (X) 
were taken by the degraded GSC units.
Dashed lines at $I_{2-20}=$0.8 and 4.5
represent the boundaries
among three different regimes (see text).
The right-side ordinate
represents the 
bolometric luminosity calculated by
assuming $D=7$ kpc, 
a bolometric correction factor in equation (\ref{equ:fbol}),
and an isotropic emission.
}
\label{fig:hid}
\end{figure}

In figure \ref{fig:hid},
we show the hardness-intensity diagrams (HIDs), 
ie. SC or HC versus 2--20 keV photon flux $\equiv I_{2-20}$,
using 2-d bin data.
As seen in figure \ref{fig:gsclc},
the periods covered by the normal GSC units,
MJD 58062-58108 and MJD 58135-58165,
are limited to the outburst decay phase,
and they have a gap from MJD 58108 to 58135.
We hence employed data taken by the degraded GSC units during the gap.
To reduce their large statistical uncertainty, 
these data were averaged over 5-d time bin.
The obtained HIDs for the SC and HC are largely represented by 
a negative intensity-hardness correlation 
when the intensity is high ($I_{2-20} \gtrsim 0.8$),
and relatively constant hardness ratios 
when the intensity is low ($I_{2-20} \lesssim 0.8$).
These features agree with those obtained from the NICER data
\citepalias{2018ApJ...863....9W}.

The two HIDs in figure \ref{fig:hid},
though grossly similar, differ in details.
In the very high-intensity region of $I_{2-20} \gtrsim 4.5$ which is
just after the outburst peak, the SC changes little with $I_{2-20}$
but the HC changes significantly.
During the intermediate region of $0.8\lesssim I_{2-20} \lesssim 4.5$,
the change of SC becomes larger, but that of HC becomes smaller than
those at $I_{2-20} \gtrsim 4.5$.  In figure \ref{fig:hid}, the
boundaries of these regions at $I_{2-20}=0.8$ and $4.5$ are marked by
dashed lines.

To clarify the source evolution during the first outburst
from MJD 58025 to 58175, we divided the time periods when 
Swift J0243.6 was observed by the normal GSC units into 8 intervals,
and named them A through H, each covering 8--10 d,
as illustrated in the top panel of figure \ref{fig:gsclc}.
These intervals have gaps
from the outburst start to MJD 58062,
and from MJD 58106 to 58134, 
for which Swift J0243.6 was observed only by the degraded GSC units.
We then
decided to use the degraded units to fill in these two gaps,
and divided them
into 5 intervals, U through Y, each of which has
a length of 8--14 d.
Table \ref{tab:outint} summarizes the start and stop time (MJD), the
employed GSC units, exposure time ($T_\mathrm{exp}$), 
and average detector area ($A_\mathrm{eff}$) for the
Swift J0243.6 direction in each interval.
Below, we employ these interval definitions.

\begin{deluxetable}{llllll}
\tablecaption{The 13 GSC data intervals for the 2017-2018 outburst phase.\label{tab:outint}}
\tablehead{
\colhead{Int.} &
\colhead{Start\tablenotemark{a}} &
\colhead{Stop\tablenotemark{a}} &
\colhead{GSC IDs} &
\colhead{$T_\mathrm{exp}$ (s)} &
\colhead{$A_\mathrm{eff}$ (cm$^2$)}
}
\startdata
U\tablenotemark{b} & 58038 & 58046 & ~~0,3,6 & ~~6115 & ~~0.981\\
V\tablenotemark{b} & 58046 & 58054 & ~~0,3,6 & 10731  & ~~1.036\\
W\tablenotemark{b} & 58054 & 58062 & ~~0,3,6 & ~~1516 & ~~1.140\\
A     & 58062 & 58070 & ~~1,4,7 & ~~2904 & ~~2.130\\
B     & 58070 & 58078 & ~~1,7   & ~~4819 & ~~3.097\\
C     & 58078 & 58086 & ~~1,7   & ~~5741 & ~~3.325\\
D     & 58086 & 58096 & ~~1,7   & ~~7102 & ~~3.287\\
E     & 58096 & 58106 & ~~1,7   & ~~5438 & ~~2.642\\
X\tablenotemark{b} & 58106 & 58120 & ~~0,3,6 & 15333  & ~~0.933\\
Y\tablenotemark{b} & 58120 & 58134 & ~~0,3,6 & ~~6962 & ~~0.852\\
F     & 58134 & 58144 & ~~1,4,7 & ~~3911 & ~~2.348\\
G     & 58144 & 58154 & ~~1,7   & ~~6535 & ~~3.243\\
H     & 58154 & 58164 & ~~1,7   & ~~7293 & ~~3.367\\
\enddata
\tablecomments{
$^a$Start and stop time in MJD.
$^b$These intervals were covered by the degraded detector units.
}
\end{deluxetable}

\subsection{Pulse profile evolution}
\label{sec:anapulse}

To study time evolution of the pulsed X-ray emission, 
we performed pulse timing analysis.  
To begin with, 
every GSC event time was converted to that
at the solar system barycenter.
Then, these barycentric times were further 
corrected for the pulsar's orbital motion,
using the binary orbital parameters shown
in table \ref{tab:orbpar}.

\begin{figure*}
\centering
\includegraphics[width=16.5cm]{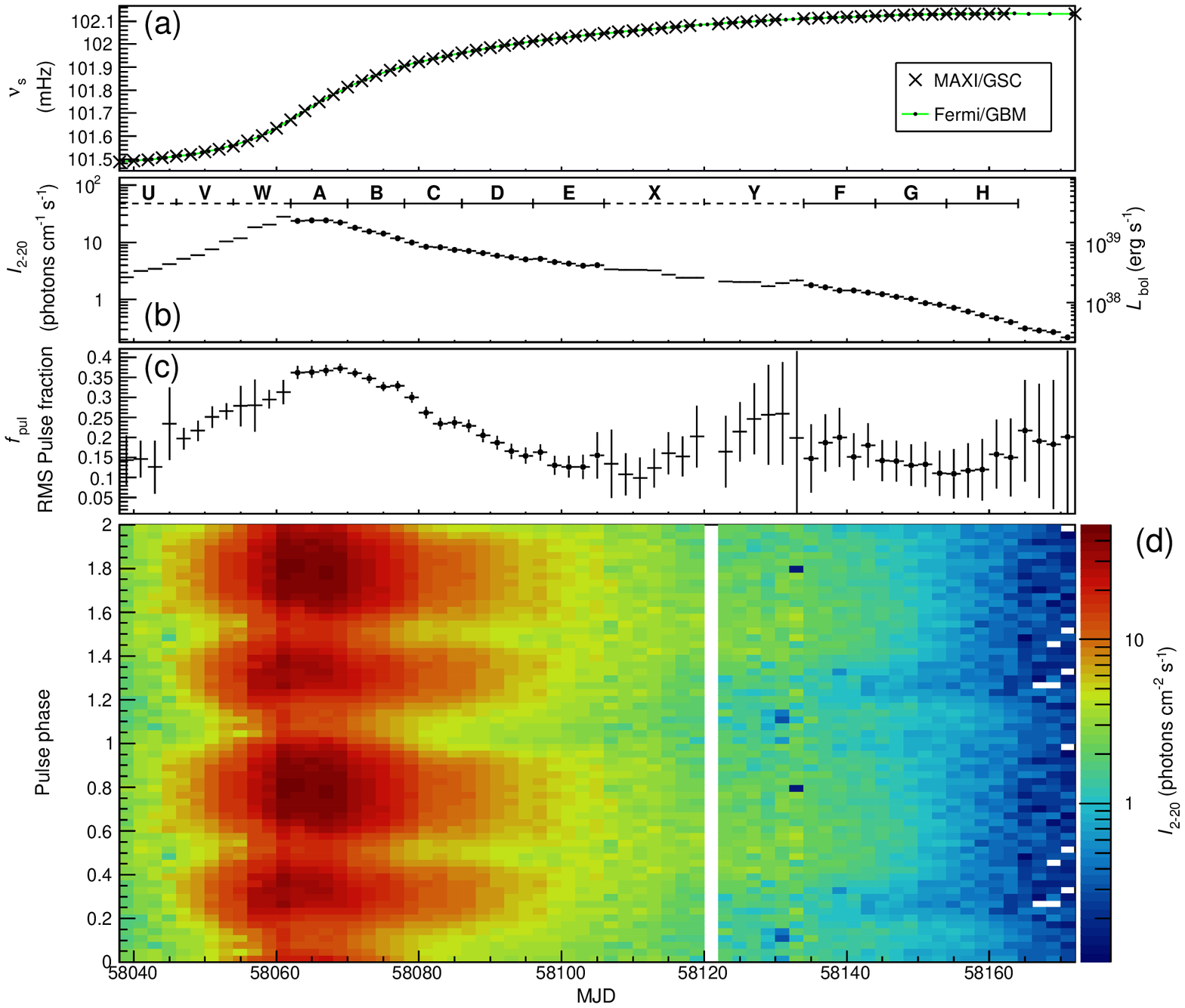} 
\caption{
  (a) The pulse frequency $\nu_{s}$ obtained with the GSC, and
  the Fermi/GBM. 
  Solid line represents the cubic spline fits. 
  Typical errors associated with the GSC and Fermi/GBM
  frequency determinations are $5\times 10^{-4}$ mHz and $5\times 10^{-6}$ mHz, respectively.
  (b) The 2-20 keV GSC photon flux $I_{2-20}$ averaged over the pulse phase. 
  The ordinate on the right-side represents the luminosity scale, same as in figure \ref{fig:hid}.
  (c) RMS pulsed fraction $f_\mathrm{pul}$ of the 2-20 keV pulse profile.
  (d) Evolution of the 2-20 keV pulse profile in color coding from MJD 58040 to 58170.
  \label{fig:perhist}
}
%
\vspace{2mm}
\centering
\includegraphics[width=8.7cm]{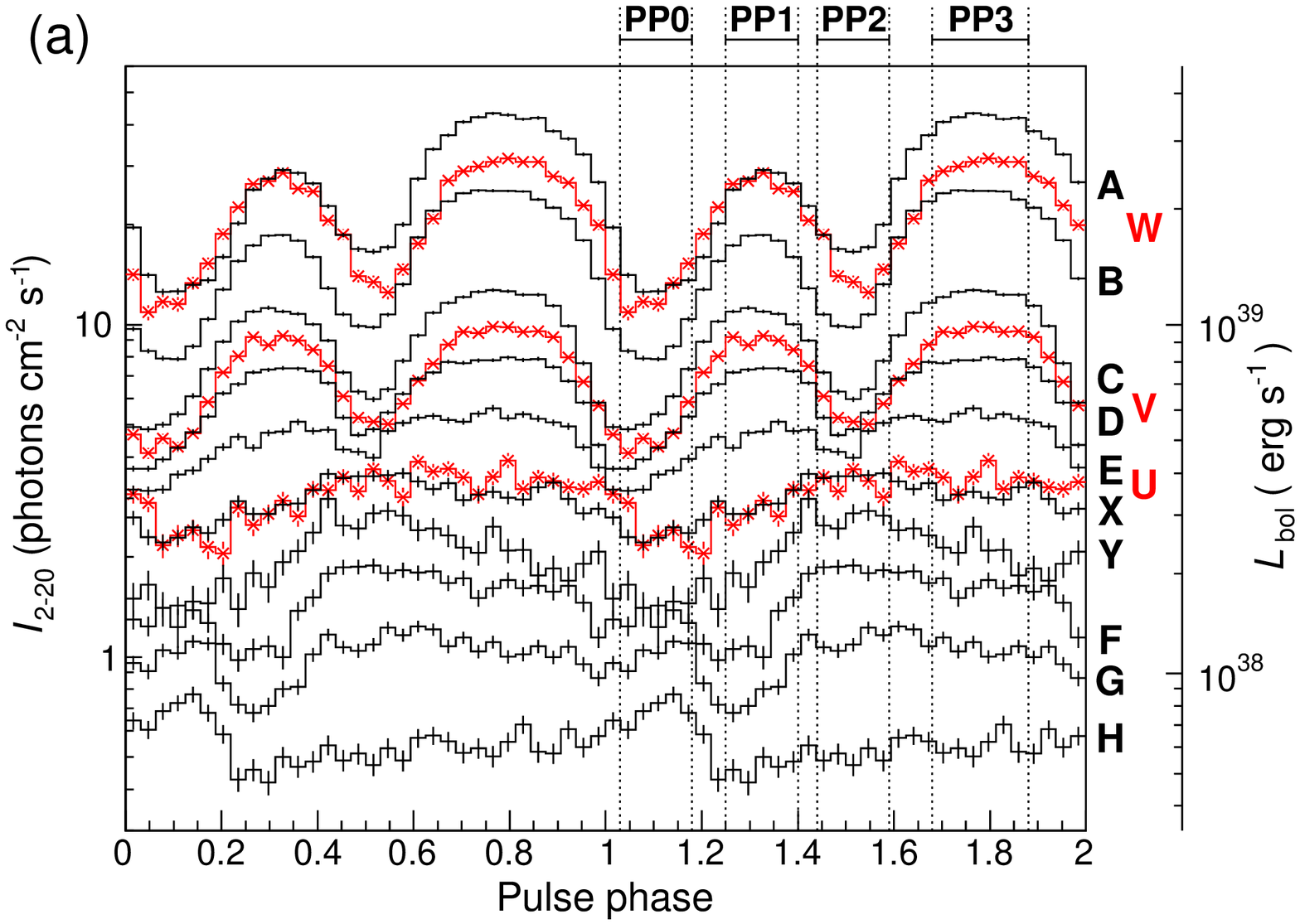} 
\hspace{3mm}
\includegraphics[width=8.7cm]{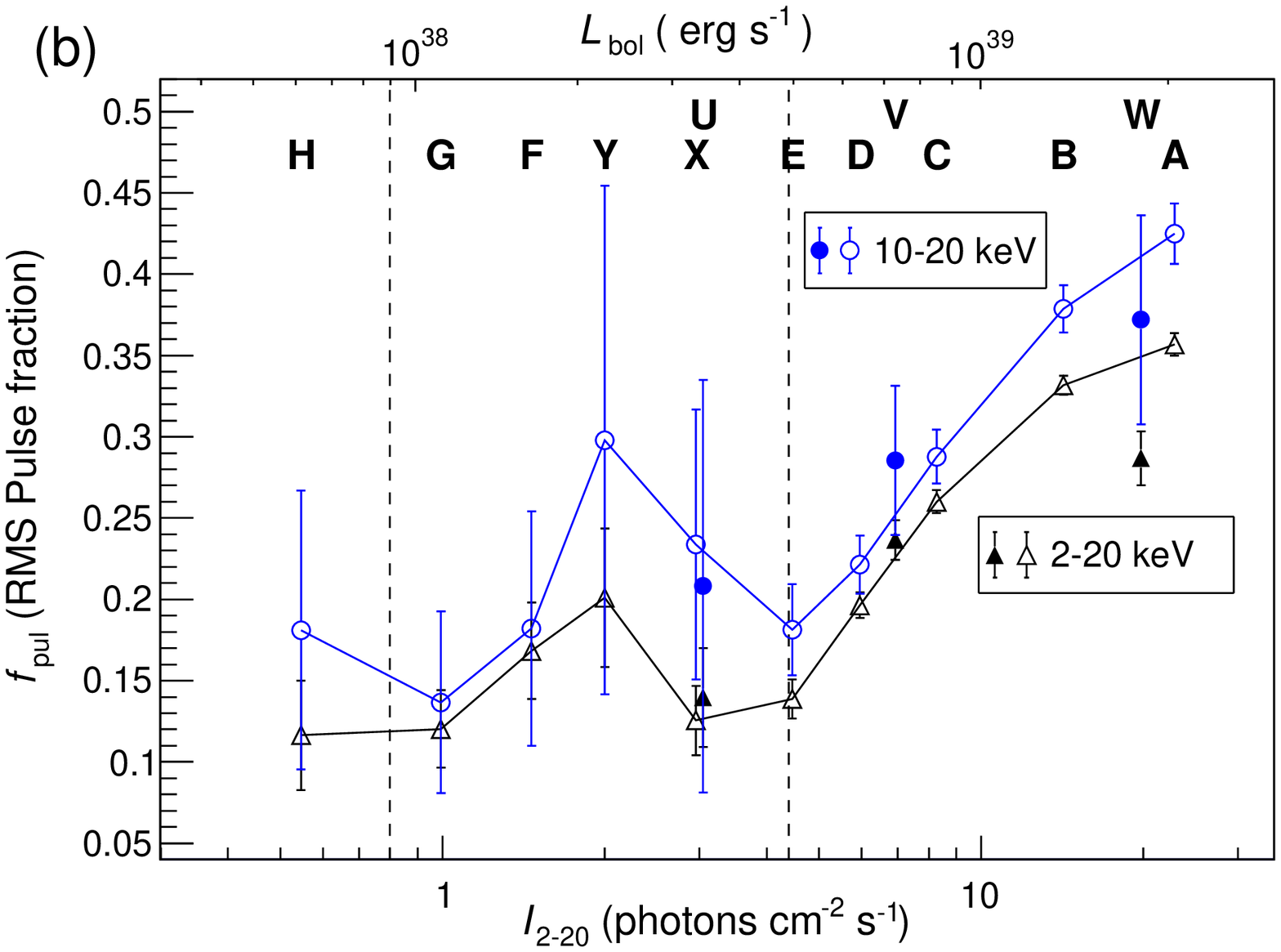} 
\caption{
(a) 2--20 keV pulse profiles for the individual intervals of
A through H, and U through Y, defined in table \ref{tab:outint}.
Profiles during the outburst rising (U, V, and W) are drawn in red
and marked by crosses (X).
Four pulse phases, PP0, PP1, PP2, and PP3,
are defined at the top.
(b) The $I_{2-20}$ dependence of $f_\mathrm{pul}$
in 2--20 keV (black circles) and 10--20 keV (blue triangles).
Data in the rising phase are plotted with filled symbols
(without lines).
Vertical dashed lines represent the two
$I_{2-20}$ boundaries characterizing the HIDs.
\label{fig:eflc}
\label{fig:pfrct}
}
\end{figure*}

We examined the coherent pulsation, first with the GSC data.
Considering the limited exposure and sparse time coverage,
the epoch-folding period search was carried out 
for every 2-d interval.
Figure \ref{fig:perhist} (a) shows 
the obtained pulse frequencies of the 2-d intervals for which
the pulsation was detected significantly,
from MJD 58038 to 58172
during the first giant outburst.
The pulse frequency of Swift J0243.6 
has also been measured  
by the Fermi/GBM 
on almost daily basis
during the X-ray active periods.  
In figure \ref{fig:perhist} (a), 
the data from the Fermi/GBM are plotted together.
We confirmed that the frequencies from the GSC data 
are all consistent with those of the Fermi/GBM 
within the errors quoted in the figure caption.

We then investigated pulse-profile evolution.
To derive phase-coherent pulse profiles considering the pulse-period changes,
we calculated a sequential pulse phase $\phi(t)$ for the event time $t$, as
\begin{equation}
\phi(t) = \int^{t}_{t_0} \nu(\tau)d\tau, 
\label{equ:pphase}
\end{equation}
where $\nu(t)$ means 
the pulse-frequency time history,
and $t_0$ is the phase-zero epoch,
i.e. $\phi(t_0)=0$.
As $\nu(t)$ to represent the observations,
we employed the daily frequencies
taken by the Fermi/GBM   
at the measured time epochs, 
because they have better accuracies than those of the GSC.
Also, $t_0$ was fixed at 58027.499066 (MJD),
which is the epoch of the first Fermi/GBM periodicity detection.
The behavior of $\nu(t)$ between adjacent data points
was estimated by a cubic spline-fit model.
In figure \ref{fig:perhist} (a), the interpolated
$\nu(t)$ model is drawn on the data.

Using equation (\ref{equ:pphase}),
we folded both the source and background light curves,
normalized them to the average detector area
for the source, and subtracted the latter from the former.
In figure \ref{fig:perhist} (d),
the pulse profiles obtained in this way
every 2-d interval
from MJD 58038 to 58172
are plotted in a 2-dimensional color image.
Figures \ref{fig:perhist} (b) and \ref{fig:perhist} (c) show
the pulse-phase-average X-ray flux and
the root-mean-square (RMS) 
pulsed fraction, $f_\mathrm{pul}$ \citepalias{2018ApJ...863....9W},
calculated from each pulse profile.
These figures reconfirm the sequential
pulse-profile change reported by \citetalias{2018ApJ...863....9W}.

Figure \ref{fig:eflc} (a) shows
the pulse profiles averaged over
the individual 8-14 d intervals of A through H, 
and U through Y,
defined in table \ref{tab:outint}.
The pulse profile changed
from a double-peak shape in the brightest phase
to a shallow single-peak one in the intermediate phase, 
and then to a dip-like feature developed in the fainter phase, 
as observed by NICER and Fermi/GBM
\citepalias{2018ApJ...863....9W}.

Figure \ref{fig:pfrct} (b) presents the $I_{2-20}$ dependence of
$f_\mathrm{pul}$, calculated from the pulse profiles in figure
\ref{fig:eflc} (a).
We also produced pulse profiles in the hard band of 10--20 keV, with
the same procedure.  The $I_{2-20}$ dependence of $f_\mathrm{pul}$ in
this band is plotted together in figure \ref{fig:pfrct} (b).
These results from the two bands confirm the NICER results
\citepalias{2018ApJ...863....9W} that the pulsed faction increase
towards higher energies.
The $f_\mathrm{pul}$ minimum at around $I_{2-20}\simeq 4.5$,
corresponding to the epoch of transition from the double-peak to the single-peak,
agrees well with the boundary of 
the two regimes in the HC-HID (right panel of figure \ref{fig:hid}).

\subsection{X-ray spectral evolution}
\label{sec:spectra}

\subsubsection{Pulse-phase-average spectra}
\label{sec:PhaseAveSpec}

The source behavior on the HIDs, 
as seen in figures \ref{fig:gsclc} and \ref{fig:hid},
suggests that the energy spectrum changed with the X-ray luminosity.  
We thus analyzed X-ray spectra taken with the GSC and
averaged over the pulse phase.
The spectral model fits were carried out on the XSPEC
software version 12.8 \citep{1996ASPC..101...17A} released as a part
of the HEASOFT software package, version 6.25.

\begin{figure*}
\centering
\includegraphics[width=8.2cm]{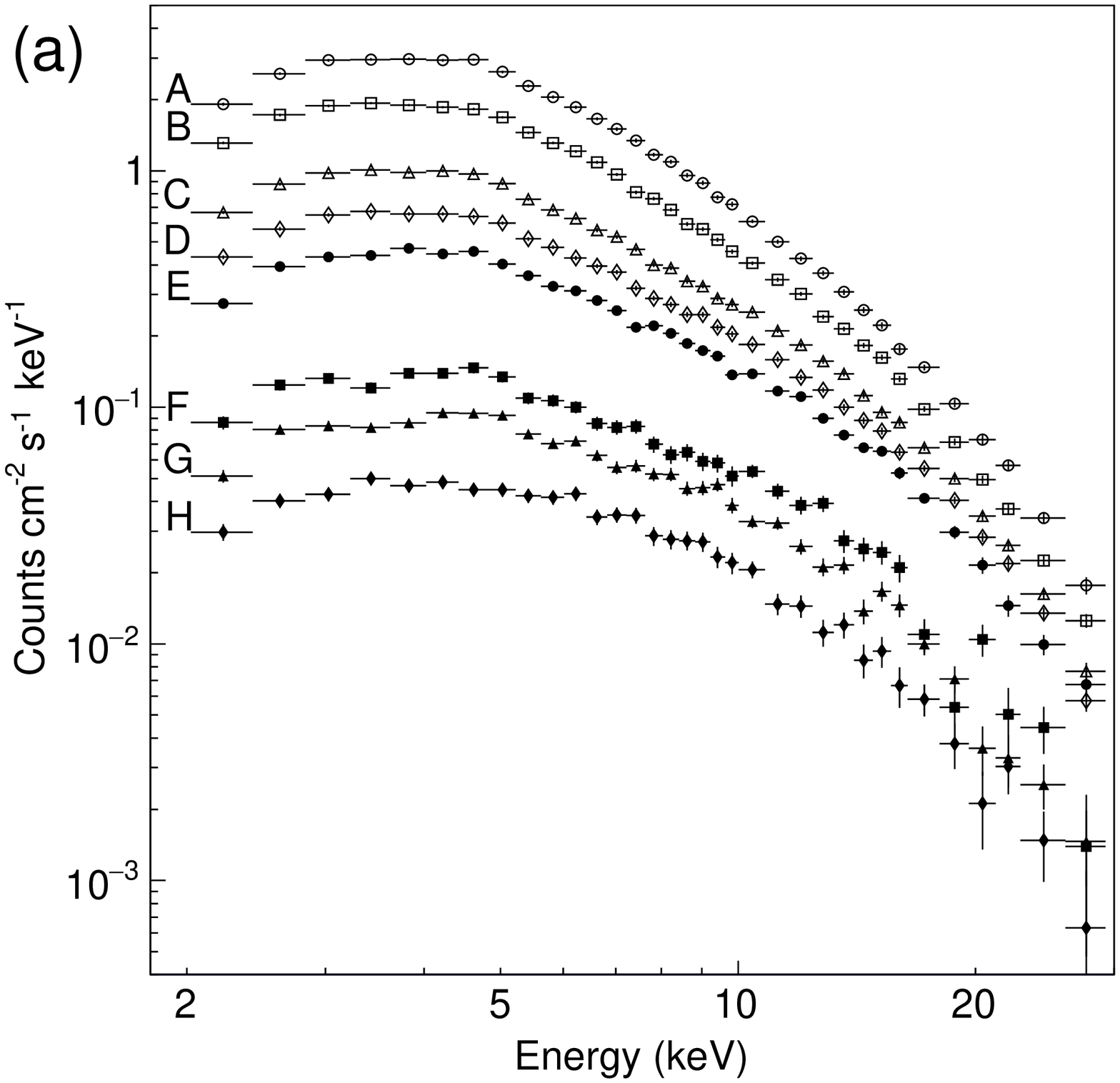} 
\hspace{4mm}
\includegraphics[width=8.2cm]{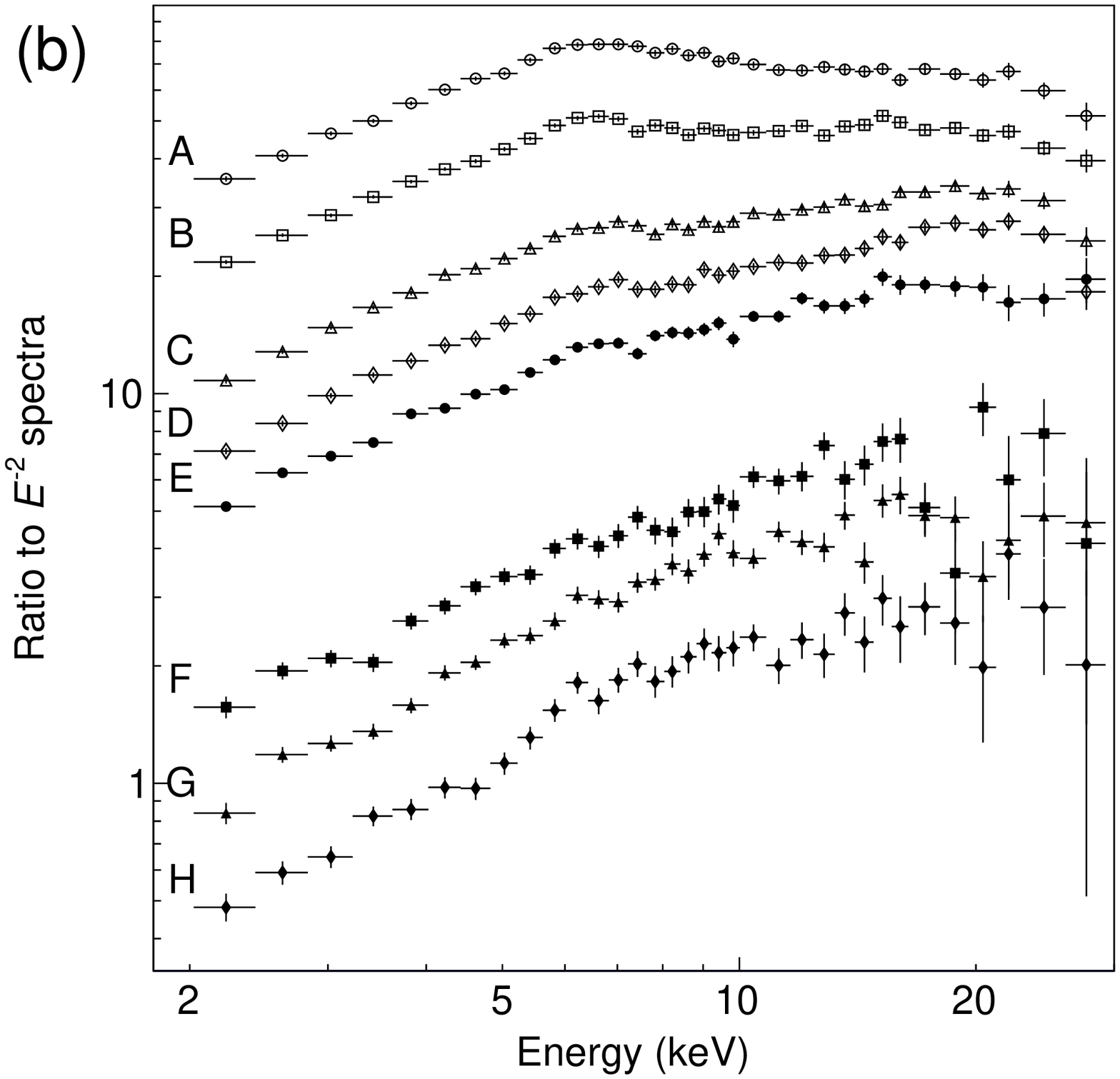}   
\caption{(a) The 2-30 keV GSC spectra for the intervals A through H,
covering the outburst decay phase.
The backgrounds have been subtracted, but
the instrument responses are inclusive.
(b) Ratios of the spectra in panel (a) to those expected for a power-law model 
$F(E)=E^{-2}$ (photons cm$^{-2}$ s$^{-1}$ keV$^{-1}$). 
\label{fig:spectra}
}
\end{figure*}

We extracted X-ray spectra 
for the 8 intervals, A through H, 
(table \ref{tab:outint}),
which were observed by the normal GSCs units. 
Figure \ref{fig:spectra} (a) shows the obtained 2--30 keV spectra, 
where the background has been subtracted as described in section \ref{sec:obs}, 
but the instrumental responses are inclusive.
To clarify the spectral evolution,
we plot in figure \ref{fig:spectra} (b) 
their ratios to the spectra expected for a power-law function 
with a photon index $\Gamma=2$, i.e. $F(E)=E^{-2}$ (photons cm$^{-2}$ s$^{-1}$ keV$^{-1}$). 
The ratios confirm the softening with the flux increase, as seen in the HIDs (figure \ref{fig:hid}).
In addition, the ratios are generally more convex than the $\Gamma=2$ power-law,
with a mild bending at 6--8 keV.
An enhancement at around 6.5 keV
is considered to 
include the iron-K line emission.

As inspired by figure \ref{fig:spectra} (b),
we fitted these spectra with a model composed of 
a high-energy-cutoff power-law (HECut)
and a Gaussian (Gaus) for the iron-K emission line.
The HECut model is represented 
by a photon index $\Gamma$, a cutoff energy
$E_\mathrm{cut}$.  a folded energy $E_\mathrm{fold}$, and a normalization factor
$A$, as a function of the photon energy $E$ as
\begin{equation} 
  F_\mathrm{HECut} = \left\{
  \begin{array}{ll}
    A E^{-\Gamma} & (E\leq E_\mathrm{cut}) \\
    A E^{-\Gamma}\exp\left(-\frac{E-E_\mathrm{cut}}{E_\mathrm{fold}} \right) & (E_\mathrm{cut}<E).
  \end{array} \right.
\end{equation}
The model has been successfully fitted to the spectra of major XBPs
\citep[e.g.][]{1983ApJ...270..711W, 2002ApJ...580..394C}.  
Because of the limited GSC energy resolution, 
we constrained
the Gaussian centroid in a 6.4--7.0 keV range,
and fixed the width at $\sigma=0.3$ keV,
referring to the spectra of the typical XBPs. 
To account for the interstellar absorption,
the continuum model was multiplied by
a photoelectric absorption factor by a medium with the Solar abundances 
\citep{2000ApJ...542..914W},
with the equivalent-hydrogen column density
fixed at the Galactic HI density in the direction,
$N_\mathrm{H}=0.7\times 10^{22}$ cm$^{-2}$
\citep{2005A&A...440..775K}.
This $N_\mathrm{H}$ value is consistent with
that determined by the NuSTAR spectrum 
in the outburst early phase
\citep{2018MNRAS.474.4432J}.
The model is hence expressed as {\tt tbabs*(powerlaw*highecut+gaussian)}
in the XSPEC terminology.

\begin{figure*}
\centering
\includegraphics[width=8.2cm]{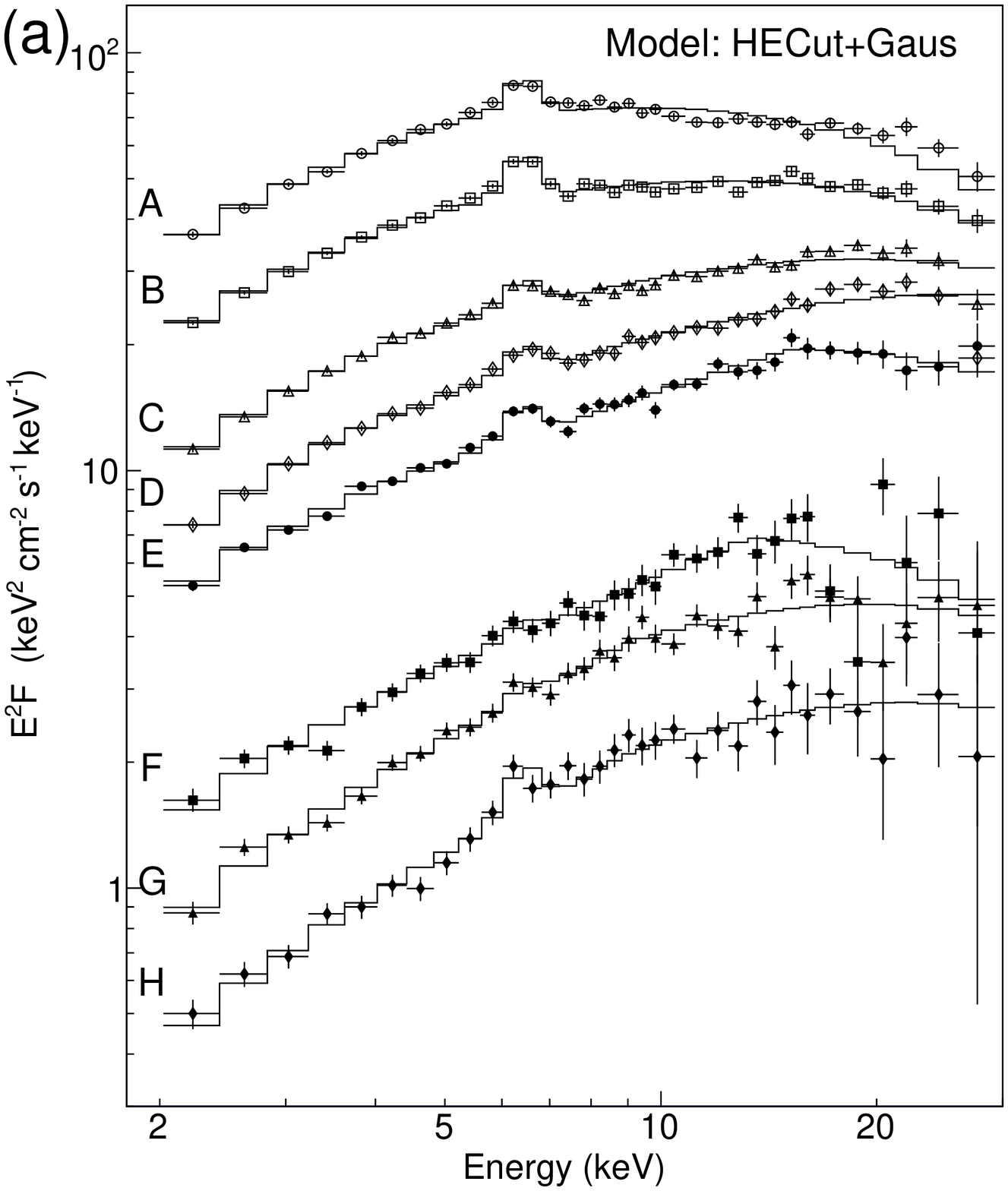} 
\hspace{4mm}
\includegraphics[width=8.2cm]{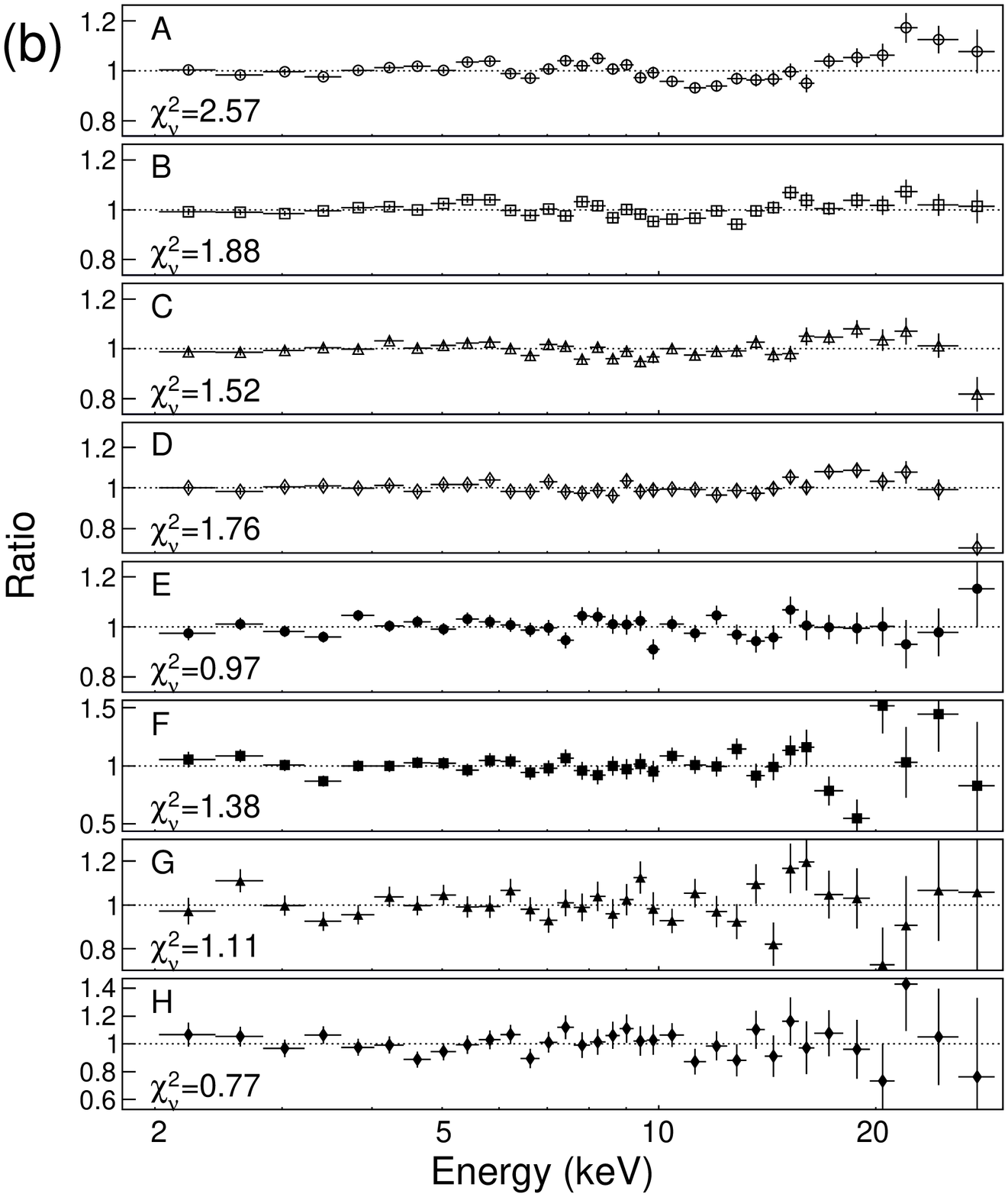} 
\caption{
(a) Unfolded $\nu F\nu$ spectra
and the best-fit HECut+Gaus models
for the intervals, A to H.
(b) Ratios of the observed spectra to the best-fit HECut+Gaus models.
}
\label{fig:eufspec}
\end{figure*}

Figure \ref{fig:eufspec} (a) shows
the unfolded $\nu F\nu$ spectra of the A through H intervals
together with their best-fit HECut$+$Gaus models,
and figure \ref{fig:eufspec} (b) shows individual data-to-model ratios. 
Table \ref{tab:specparam} summarizes the best-fit model parameters
which include the absorption-corrected 0.5--60 keV flux,
$F_{0.5-60}$, considered to approximate the bolometric flux.
The value of $F_{0.5-60}= 38$ erg cm$^{-2}$ s$^{-1}$
in the interval A corresponds to
the bolometric luminosity $L_\mathrm{bol}= 2.2\times 10^{39}$ erg s$^{-1}$
assuming an isotropic emission and $D=7$ kpc.
Although the HECut$+$Gaus model largely reproduced the data, 
the data-to-model ratios are not always consistent with 1.
The discrepancies are evident
in higher-luminosity intervals and 
in energies $\gtrsim 6$ keV.
The 
$\chi^2_\nu$ values indicate that the fits 
are not acceptable within the 95 \% confidence limits
in the first half of the observation, the intervals A through D,
but those of the the second half, E to H, 
are acceptable.

\begin{deluxetable*}{cllllllllll}
\tablecaption{The best-fit spectral parameters with the HECut+Gaus and NPEX+Gaus models.\label{tab:specparam}}
\tablehead{
& \multicolumn{10}{c}{Model: HECut + Gaus}\\
& \colhead{$A$}
& \colhead{$\Gamma$} 
& \colhead{$E_\mathrm{cut}$}
& \colhead{$E_\mathrm{fold}$} 
& \colhead{$E_\mathrm{Fe}$$^{a}$}
& \colhead{$EW_\mathrm{Fe}$$^{b}$} 
& \colhead{$I_\mathrm{2-20}$$^{c}$}
& \colhead{$F_\mathrm{0.5-60}$$^{d}$} 
& \colhead{$f_\mathrm{bol}$$^{e}$} 
& \colhead{$\chi^2_\nu (\nu)$} \\
\colhead{Int.} & & 
& \colhead{(keV)}
& \colhead{(keV)}
& \colhead{(keV)} 
& \colhead{(eV)}
}
\startdata
A & $32^{*}$ & $~1.52^{*}$ & $~5.2^{*}$ & $19^{*}$ & $~6.4^{*}$ & $170^{*}$ & $25.3^{*}$ & $37.9^{*}$ & $~1.50^{*}$ & 2.66 (28) \\
B & $20^{+1}_{-1}$ & $~1.50^{+0.03}_{-0.03}$ & $~4.8^{+0.5}_{-0.5}$ & $24^{+2}_{-2}$ & $~6.4^{+0.03}_{-0.00}$ & $210^{+40}_{-30}$ & $16.2^{+0.1}_{-0.1}$ & $26.0^{+0.5}_{-0.4}$ & $~1.61^{+0.04}_{-0.04}$ & 1.95 (28) \\
C & $~9.1^{+0.4}_{-0.2}$ & $~1.41^{+0.04}_{-0.04}$ & $~4.2^{+0.8}_{-0.7}$ & $32^{+5}_{-4}$ & $~6.4^{+0.11}_{-0.00}$ & $120^{+40}_{-40}$ & $~8.74^{+0.06}_{-0.06}$ & $15.9^{+0.4}_{-0.4}$ & $~1.82^{+0.06}_{-0.06}$ & 1.58 (28) \\
D & $~5.6^{+0.4}_{-0.3}$ & $~1.35^{+0.05}_{-0.05}$ & $~3.8^{+1.7}_{-1.3}$ & $40^{+11}_{-7}$ & $~6.4^{+0.14}_{-0.00}$ & $120^{+40}_{-30}$ & $~6.05^{+0.04}_{-0.04}$ & $12.1^{+0.4}_{-0.4}$ & $~2.01^{+0.08}_{-0.08}$ & 1.82 (28) \\
E & $~4.5^{+0.17}_{-0.16}$ & $~1.46^{+0.02}_{-0.02}$ & $15^{+3}_{-4}$ & $28^{+27}_{-11}$ & $~6.4^{+0.12}_{-0.00}$ & $140^{+50}_{-40}$ & $~4.36^{+0.05}_{-0.05}$ & $~8.4^{+0.8}_{-0.7}$ & $~1.92^{+0.21}_{-0.17}$ & 1.01 (28) \\
F & $~1.11^{+0.09}_{-0.09}$ & $~1.29^{+0.05}_{-0.05}$ & $12.7^{+2.4}_{-2.5}$ & $18^{+18}_{-8}$ & $~6.4^{+0.6}_{-0.0}$ & $40^{+90}_{-40}$ & $~1.40^{+0.03}_{-0.03}$ & $~2.5^{+0.5}_{-0.3}$ & $~1.82^{+0.39}_{-0.27}$ & 1.44 (28) \\
G & $~0.57^{+0.05}_{-0.05}$ & $~1.12^{+0.06}_{-0.07}$ & $~9.0^{+2.0}_{-2.1}$ & $22^{+9}_{-5}$ & $~6.4^{+0.6}_{-0.0}$ & $40^{+120}_{-40}$ & $~0.93^{+0.02}_{-0.02}$ & $~1.90^{+0.22}_{-0.18}$ & $~2.03^{+0.27}_{-0.23}$ & 1.15 (28) \\
H & $~0.29^{+0.04}_{-0.03}$ & $~1.09^{+0.08}_{-0.08}$ & $~8.9^{+2.4}_{-2.0}$ & $24^{+20}_{-8}$ & $~6.4^{+0.6}_{-0.0}$ & $210^{+140}_{-130}$ & $~0.51^{+0.01}_{-0.01}$ & $~1.09^{+0.21}_{-0.16}$ & $~2.13^{+0.47}_{-0.36}$ & 0.80 (28) \\
\hline
\hline
& \multicolumn{10}{c}{Model: NPEX + Gauss}\\
& \colhead{$A_{1}$} 
& \colhead{$\Gamma_1$}
& \colhead{$A_{2}(\times 10^3)$}  
& \colhead{$E_\mathrm{fold}$}
& \colhead{ $E_\mathrm{Fe}$$^{a}$}
& \colhead{ $EW_\mathrm{Fe}$$^{b}$} 
& \colhead{$I_\mathrm{2-20}$$^{c}$}
& \colhead{$F_\mathrm{0.5-60}$$^{d}$} 
& \colhead{$f_\mathrm{bol}$$^{e}$}
& \colhead{$\chi^2_\nu (\nu)$} \\
\colhead{Int.} & & & 
& \colhead{(keV)}
& \colhead{(keV)} 
& \colhead{(eV)}
& \\
\hline
A & $29^{*}$ & $~1.04^{*}$ & $~3.2^{*}$ & $~7.2^{*}$ & $~6.4^{*}$ & $180^{*}$ & $25.3^{*}$ & $37.4^{*}$ & $~1.48^{*}$ & 2.25 (28) \\
B & $18^{+1}_{-0}$ & $~0.96^{+0.06}_{-0.04}$ & $~5.3^{+2.6}_{-2.3}$ & $~6.1^{+0.9}_{-0.6}$ & $~6.4^{*}$ & $210^{+30}_{-30}$ & $16.2^{+0.1}_{-0.1}$ & $24.5^{+0.9}_{-0.6}$ & $~1.51^{+0.06}_{-0.05}$ & 1.55 (29) \\
C & $~8.5^{+0.4}_{-0.3}$ & $~0.89^{+0.03}_{-0.03}$ & $~5.1^{+1.6}_{-1.5}$ & $~5.8^{+0.5}_{-0.4}$ & $~6.4^{+0.15}_{-0.00}$ & $130^{+40}_{-30}$ & $~8.78^{+0.06}_{-0.06}$ & $14.3^{+0.5}_{-0.4}$ & $~1.63^{+0.06}_{-0.05}$ & 0.97 (28) \\
D & $~5.4^{+0.3}_{-0.2}$ & $~0.86^{+0.03}_{-0.03}$ & $~4.5^{+1.3}_{-1.2}$ & $~5.8^{+0.5}_{-0.4}$ & $~6.4^{+0.18}_{-0.00}$ & $140^{+40}_{-30}$ & $~6.10^{+0.05}_{-0.05}$ & $10.5^{+0.4}_{-0.3}$ & $~1.72^{+0.07}_{-0.06}$ & 1.24 (28) \\
E & $~4.0^{+0.4}_{-0.3}$ & $~0.86^{+0.05}_{-0.05}$ & $~5.1^{+1.9}_{-1.7}$ & $~5.2^{+0.6}_{-0.4}$ & $~6.4^{+0.16}_{-0.00}$ & $140^{+70}_{-40}$ & $~4.37^{+0.05}_{-0.05}$ & $~7.4^{+0.4}_{-0.3}$ & $~1.68^{+0.11}_{-0.08}$ & 1.09 (28) \\
F & $~1.32^{+0.32}_{-0.26}$ & $~0.84^{+0.17}_{-0.15}$ & $~4.7^{+2.5}_{-2.0}$ & $~4.1^{+0.6}_{-0.4}$ & $~6.4^{+0.6}_{-0.0}$ & $90^{+150}_{-90}$ & $~1.41^{+0.03}_{-0.03}$ & $~2.2^{+0.2}_{-0.1}$ & $~1.58^{+0.17}_{-0.13}$ & 1.51 (28) \\
G & $~0.49^{+0.06}_{-0.05}$ & $~0.84^{+0.13}_{-0.15}$ & $~0.00^{+0.04}_{-0.00}$ & $17^{+7}_{-6}$ & $~6.4^{+0.6}_{-0.0}$ & $20^{+140}_{-20}$ & $~0.93^{+0.02}_{-0.02}$ & $~1.90^{+0.32}_{-0.18}$ & $~2.03^{+0.38}_{-0.22}$ & 1.29 (28) \\
H & $~0.36^{+0.14}_{-0.10}$ & $~0.71^{+0.25}_{-0.22}$ & $~2.0^{+1.5}_{-2.0}$ & $~4.0^{+1.1}_{-0.5}$ & $~6.4^{+0.5}_{-0.0}$ & $280^{+160}_{-150}$ & $~0.52^{+0.01}_{-0.01}$ & $~0.83^{+0.12}_{-0.07}$ & $~1.61^{+0.28}_{-0.18}$ & 0.93 (28) \\ 
\enddata
\tablecomments{
{$^*$}{Errors are given with the 90\% limits of statistical uncertainy if the fits are within the acceptable level ($\chi^2_\nu<2$).}\\
$^{a}$Centroid and
$^{b}$equivalent width of iron-K line. \\
$^{c}$Photon flux in 2-20 keV in photon cm$^{-2}$ s$^{-1}$.\\
$^{d}$Absorption-corrected flux in 0.5-60 keV in $10^{-8}$ erg cm$^{-2}$ s$^{-1}$.\\
$^{e}$Ratio of $I_{2-20}$ to $F_{0.5-60}$ in $10^{-8}$ erg photon$^{-1}$.
}
\end{deluxetable*}

\begin{figure*}
\centering
\includegraphics[width=8.4cm]{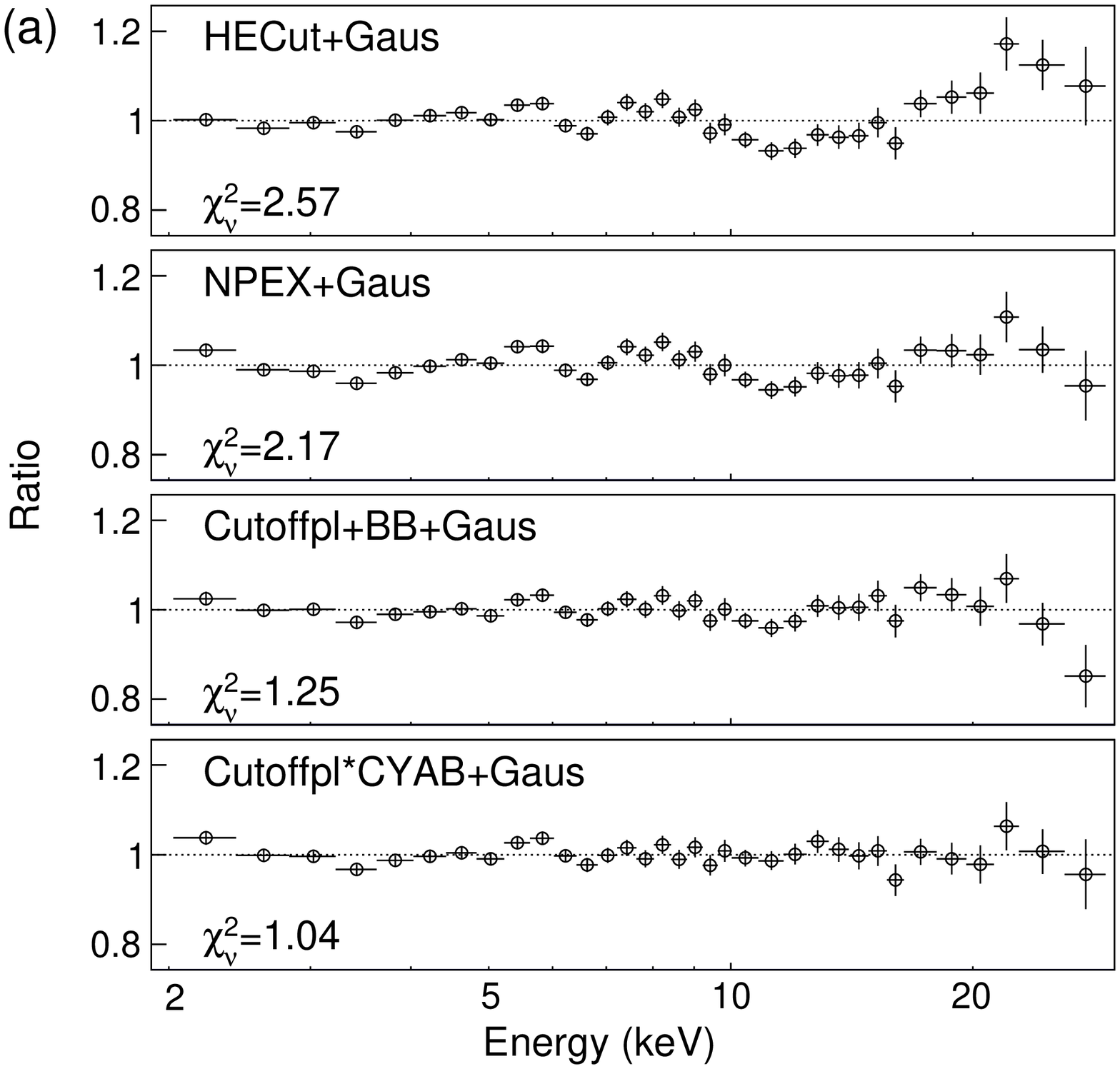}
\hspace{4mm}
\includegraphics[width=8.4cm]{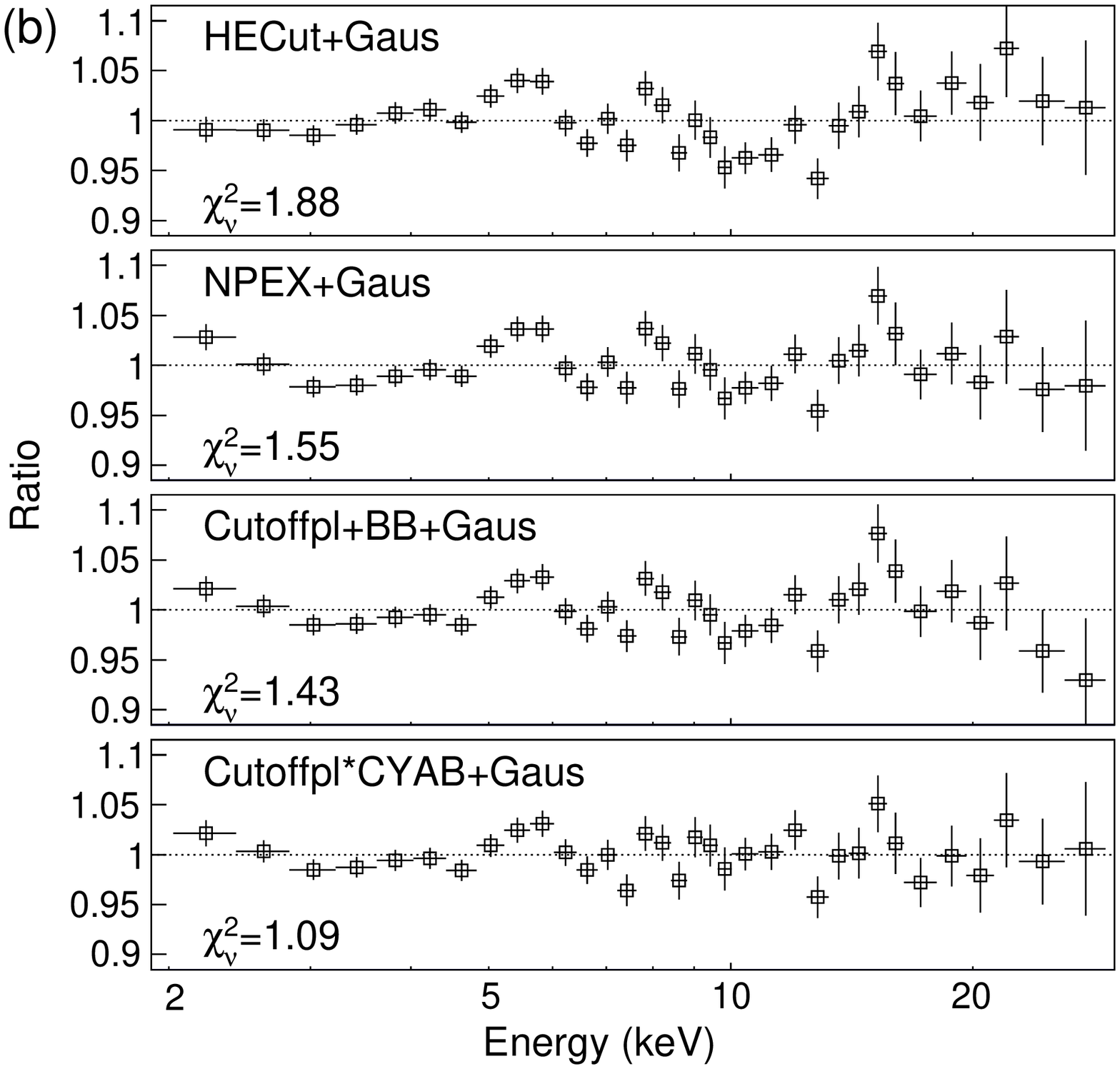}
\caption{
  Ratios of the observed spectra for intervals A (panel a) and
  B (panel b) to the best-fit models with HECut+Gaus, NPEX+Gaus,
  Cutoffpl+BB+Gaus, and Cutoffpl*CYAB+Gaus, from the top to bottom
  panels.  The best-fit $\chi^2_\nu$ value (in tables
  \ref{tab:specparam} and \ref{tab:specparam2}) are presented in each
  panel.  \label{fig:ratio2}
}
\end{figure*}

We then examined another continuum model,
NPEX \citep[Negative and Positive power laws with a common EXponential cutoff,][]{1998AdSpR..22..987M},
which has been used in the study of XBPs
often more successfully than the HECut model.
The NPEX model is represented by
\begin{equation} 
F_\mathrm{NPEX} = 
\left(A_1 E^{-\Gamma_1} + A_2 E^{+\Gamma_2} \right)
\exp\left(-\frac{E}{E_\mathrm{fold}} \right),
\end{equation}
with five parameters, $\Gamma_1$, $\Gamma_2$, $A_1$, $A_2$, and $E_\mathrm{fold}$. 
We fixed $\Gamma_2(>0)$ at the typical value of 2.0 \citep{1998AdSpR..22..987M}.
The best-fit NPEX+Gaus model parameters are listed in 
table \ref{tab:specparam}.
The fits have been improved, particularly when the source 
is luminous. 
However, the $\chi^2_\nu$ values are 
still unacceptable in the intervals A and B. 
In figure \ref{fig:ratio2}, the data-to-model ratios
are presented.
Above 10 keV, 
they still exhibit a feature that is similar to those in the
HECut+Gaus model.

This characteristic excess feature has already been noticed in the
NuSTAR and the NICER data
\citep{2019ApJ...873...19T,2019ApJ...885...18J}.  There, it was
considered as a ``broad iron line'', and thus fitted with a Gaussian
with $\sigma\sim1.5$ keV.
%
We hence attempted to fit the GSC spectra 
with a model consisting of an NPEX continuum, 
plus three Gaussians representing three lines at 
fixed energies of 6.4, 6.7 and 7.0 keV.
The 6.4 keV line was allowed to take a free width,
whereas the other two were assumed to be narrow.
The fit was acceptable with $\chi^2_\nu=1.07$ (26 degree of freedom).
The spectrum in the interval A (= outburst peak)
gave the 6.4 keV width of $\sigma= 1.27^{+0.27}_{-0.29}$ keV 
and the equivalent width of $EW_\mathrm{Fe}=0.54^{+0.20}_{-0.17}$ keV,
which are consistent with those measured with 
NICER and NuSTAR spectra \citep{2019ApJ...873...19T,2019ApJ...885...18J}.

Although the excess feature in the GSC spectra 
can be thus interpreted as a broad iron line, 
its origin is not necessarily clear  \citep[][also see later discussion]{2019ApJ...885...18J}.
Therefore, other interpretations should be explored.
%
The characteristic excess also reminds us of the ``10 keV feature'' that has been
observed in several XBPs \citep[e.g.][]{2002ApJ...580..394C}, and
interpreted either as a bump or an absorption on the cutoff power-law
continuum \citep[][]{2008A&A...491..833K}.
In the bump case, it can be fitted with
a broad Gaussian \citep[e.g.][]{2013A&A...551A...6M,2013A&A...551A...1R}
or a blackbody (hereafter BB) \citep{1999MNRAS.302..700R}.
In the absorption case, 
it can look like a cyclotron-resonance absorption 
\citep[CYAB;][]{1990Natur.346..250M}
We hence repeated the model fits by 
incorporating either a BB (bump case) or a CYAB model (absorption case)
to the HECut or the NPEX continuum.

\begin{deluxetable*}{clllllllllll}
\tablecaption{The best-fit spectral parameters with the Cutoffpl+BB+Gaus and Cutoffpl*CYAB+Gaus models.
\label{tab:specparam2}}
\tablehead{
& \multicolumn{11}{c}{Model: Cutoffpl + BB + Gaus}\\
& \colhead{$A$}
& \colhead{$\Gamma$}
& \colhead{$E_\mathrm{fold}$}
& \colhead{$kT_\mathrm{BB}$$^a$}
& \colhead{$R_\mathrm{BB}$$^b$}
& \colhead{---}
& \colhead{$E_\mathrm{Fe}$}
& \colhead{$EW_\mathrm{Fe}$} 
& \colhead{$F_\mathrm{0.5-60}$} 
& \colhead{$f_\mathrm{bol}$}
& \colhead{$\chi^2_\nu (\nu)$} \\
\colhead{Int.} & & 
& \colhead{(keV)}
& \colhead{(keV)}
& \colhead{(km)} & 
& \colhead{(keV)} 
& \colhead{(eV)} 
}
\startdata
A & $32^{+2}_{-2}$ & $~1.63^{+0.10}_{-0.10}$ & $49^{+70}_{-18}$ & $~1.61^{+0.09}_{-0.10}$ & $17^{+2}_{-2}$ & \colhead{---} & $~6.4^{+0.12}_{-0.00}$ & $120^{+50}_{-40}$ & $41.3^{+1.7}_{-1.6}$ & $1.64^{+0.08}_{-0.07}$ & 1.29 (27) \\
B & $18^{+1}_{-1}$ & $~1.43^{+0.07}_{-0.06}$ & $27^{+8}_{-5}$ & $~1.34^{+0.14}_{-0.14}$ & $14.0^{+3.2}_{-2.5}$ & \colhead{---} & $~6.4^{+0.04}_{-0.00}$ & $190^{+50}_{-30}$ & $26.4^{+0.8}_{-0.7}$ & $1.63^{+0.06}_{-0.05}$ & 1.48 (27) \\
C & $~7.8^{+0.8}_{-1.1}$ & $~1.28^{+0.08}_{-0.10}$ & $28^{+8}_{-6}$ & $~1.08^{+0.21}_{-0.16}$ & $13.6^{+7.1}_{-4.6}$ & \colhead{---} & $~6.4^{+0.18}_{-0.00}$ & $120^{+30}_{-50}$ & $15.8^{+0.6}_{-0.6}$ & $1.81^{+0.08}_{-0.08}$ & 1.27 (27) \\
D & $~4.9^{+0.8}_{-1.0}$ & $~1.23^{+0.10}_{-0.15}$ & $31^{+12}_{-8}$ & $~0.94^{+0.36}_{-0.16}$ & $12.3^{+10.1}_{-7.4}$ & \colhead{---} & $~6.4^{+0.17}_{-0.00}$ & $130^{+60}_{-40}$ & $11.9^{+0.5}_{-0.6}$ & $1.96^{+0.10}_{-0.11}$ & 1.74 (27) \\
\hline
\hline
& \multicolumn{11}{c}{Model: Cutoffpl*CYAB + Gaus}\\
& \colhead{$A$} 
& \colhead{$\Gamma$}
& \colhead{$E_\mathrm{fold}$}
& \colhead{$E_\mathrm{a}$$^c$}
& \colhead{$W_\mathrm{a}$$^d$}
& \colhead{$D_\mathrm{a}$$^e$}
& \colhead{$E_\mathrm{Fe}$} 
& \colhead{$EW_\mathrm{Fe}$} 
& \colhead{$F_\mathrm{0.5-60}$}
& \colhead{$f_\mathrm{bol}$} 
& \colhead{$\chi^2_\nu (\nu)$} \\
\colhead{Int.} & & 
& \colhead{(keV)} 
& \colhead{(keV)} 
& \colhead{(keV)} & 
& \colhead{(keV)} 
& \colhead{(eV)} 
& \\
\hline
A & $28^{+1}_{-1}$ & $~1.22^{+0.05}_{-0.05}$ & $15^{+2}_{-1}$ & $10.9^{+0.5}_{-0.7}$ & $~3.7^{+1.6}_{-1.1}$ & $~0.22^{+0.04}_{-0.04}$ & $~6.4^{+0.09}_{-0.00}$ & $110^{+40}_{-30}$ & $37.5^{+0.3}_{-0.3}$ & $1.48^{+0.02}_{-0.02}$ & 1.08 (26) \\
B & $18^{+1}_{-1}$ & $~1.24^{+0.05}_{-0.05}$ & $16^{+1}_{-1}$ & $~9.5^{+0.6}_{-1.0}$ & $~3.2^{+1.5}_{-1.1}$ & $~0.14^{+0.03}_{-0.03}$ & $~6.4^{+0.04}_{-0.00}$ & $160^{+30}_{-30}$ & $25.1^{+0.2}_{-0.2}$ & $1.55^{+0.02}_{-0.02}$ & 1.13 (26) \\
C & $~8.4^{+0.5}_{-0.6}$ & $~1.20^{+0.06}_{-0.09}$ & $21^{+3}_{-3}$ & $~8.5^{+0.9}_{-2.2}$ & $~3.2^{+2.1}_{-1.4}$ & $~0.12^{+0.04}_{-0.03}$ & $~6.4^{+0.3}_{-0.0}$ & $80^{+40}_{-40}$ & $15.3^{+0.2}_{-0.2}$ & $1.75^{+0.04}_{-0.03}$ & 0.98 (26) \\
D & $~5.4^{+0.3}_{-0.9}$ & $~1.21^{+0.07}_{-0.36}$ & $27^{+6}_{-9}$ & $~8.2^{+1.7}_{-8.2}$ & $~3.2^{+2.8}_{-1.5}$ & $~0.08^{+0.06}_{-0.04}$ & $~6.4^{+0.3}_{-0.0}$ & $100^{+80}_{-50}$ & $11.8^{+0.2}_{-0.2}$ & $1.94^{+0.05}_{-0.04}$ & 1.63 (26) \\
\enddata
\tablecomments{
$^*$Errors are given with the 90\% limits of statistical uncertainy if the fits are within the acceptable level ($\chi^2_\nu<2$).\\
$^a$Temperature and $^b$radius of BB emission assuming the distance $D=7$ kpc.\\
$^c$Cyclotron-resonance energy, $^d$width, and $^e$depth in CYAB model.
}
\end{deluxetable*}

\begin{figure*}
\centering
\includegraphics[width=8.7cm]{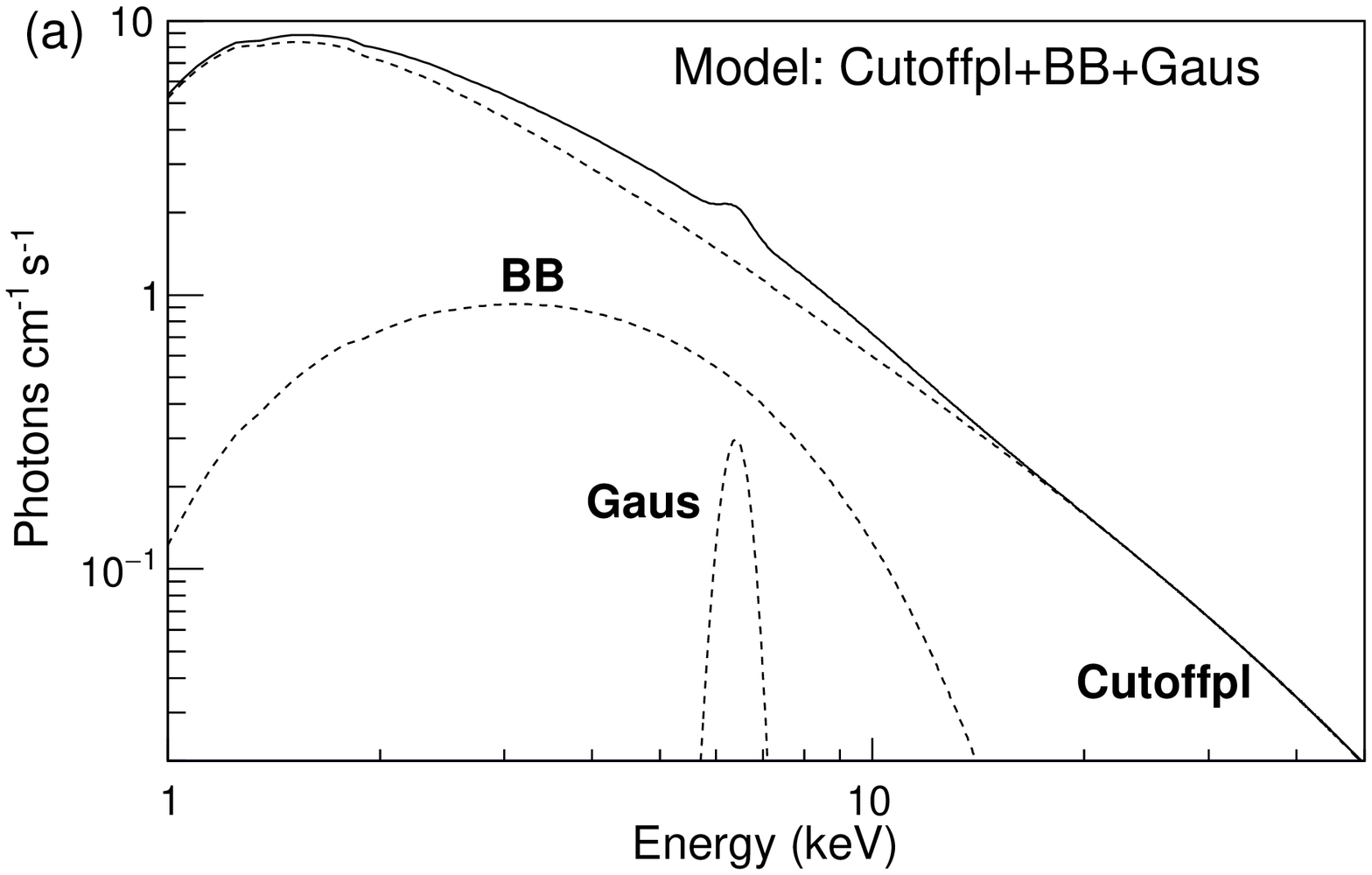}
\hspace{3mm}
\includegraphics[width=8.7cm]{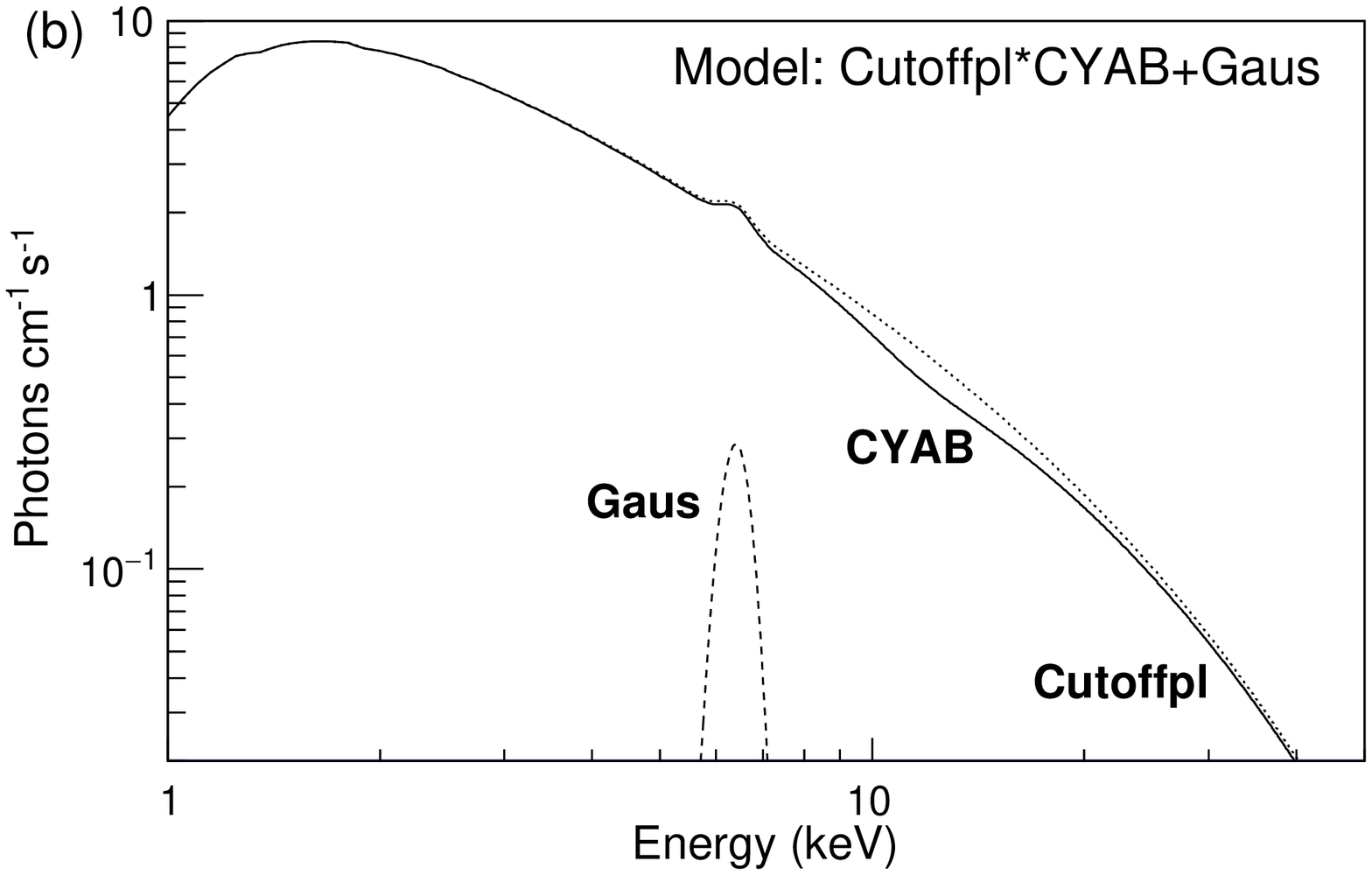}
\caption{
Two best-fit models for the interval-A spectrum,
represented by
(a) Cutoffpl+BB+Gaus and (b) Cutoffpl*CYAB+Gaus.
\label{fig:fitmodel}
}
\end{figure*}

Table \ref{tab:specparam2} summarizes
the best-fit parameters of these models for the A, B, C, and D spectra.
Becuse $E_\mathrm{cut}$ in HECut or $A_2$ in NPEX was consistent with
0, the continuum in both models can be replaced by a simple cutoff
power law (Cutoffpl) as
$F_\mathrm{Cutoffpl}=A\exp(-E/E_\mathrm{fold})$.  Therefore, the
results are given in simple model forms as Cutoffppl+BB+Gaus and
Cutoffpl*CYAB+Gaus.
Figure \ref{fig:ratio2} compares data-to-model ratios of the intervals
A and B, when using the modeling of (1) HECut+Gaus, (2) NPEX+Gaus, (3)
Cutoffpl+BB+Gaus, and (4) Cutoffpl*CYAB+Gaus.  The fits are
significantly improved by adding the BB or CYAB component.  
In the first two models of HECut+Gaus and NPEX+Gaus,
the ratios show a dip-like structure at 6.4 keV,
because the broad excess feature was fitted with a narrow Gaussian line.
It was reduced in the latter two models.
%
Figure \ref{fig:fitmodel} shows the implied Cutoffpl+BB+Gaus and
Cutoffpl*CYAB+Gaus models that give the best fits to the interval-A
spectrum.

While the latter two models are in the acceptable levels,
their data-to-model ratios in figure \ref{fig:ratio2}
still seem to have a small ($\lesssim 3$ \%) structure 
at around 5 keV.
This is considered partly due to the systematic errors on the GSC 
response function,
associated with the Xe-L edge at 4.8 keV \citep{Mihara_pasj2011}.
We confirmed that the model-fit results did not change significantly even if its energy range (4.5-5.5 keV) 
was masked.
%
%
%

\begin{figure}
\centering
\includegraphics[width=8.2cm]{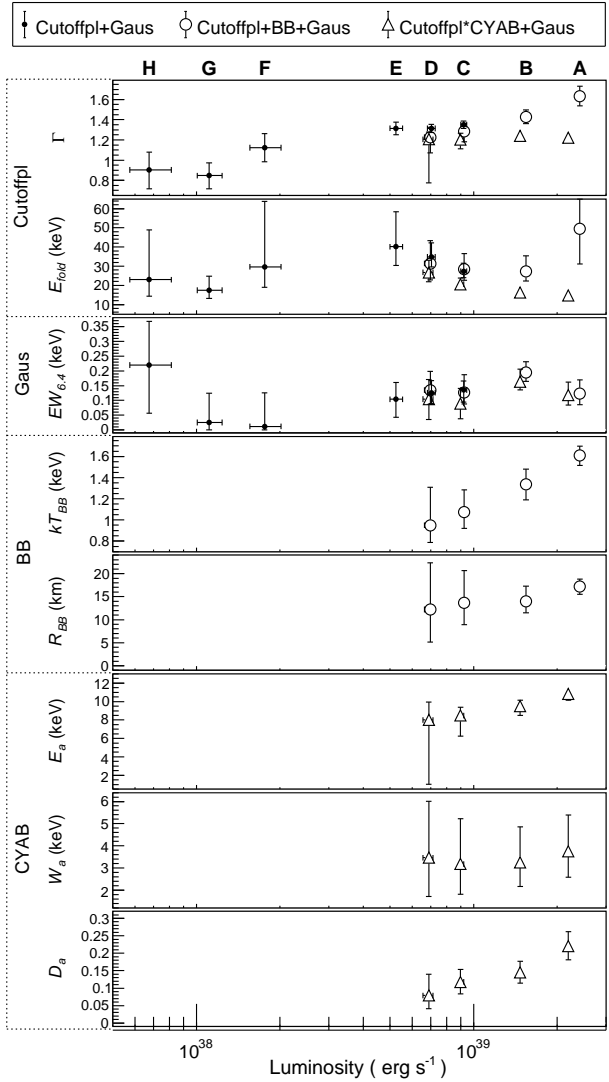} 
\caption{
Dependence of the spectral parameters on the X-ray luminosity,
obtained from the model fits to the interval A to H spectra
with {Cutoffpl+Gaus}, 
{Cutoffpl+BB+Gaus}, and 
{Cutoffpl*CYAB+Gaus}. 
Error bars represent 90\% confidence limits of the statistical uncertainties.
Only those fits that are acceptable within the 90\% confidence are shown.
\label{fig:specparam_evolve}
}
\end{figure}

To visualize the spectral-parameter evolutions, figure
\ref{fig:specparam_evolve} summarizes these best-fit parameters
against the X-ray luminosity, where plotted are results with the
HECut+Gaus, Cutoffpl+BB+Gaus, and Cutoffpl*CYAB+Gaus fits, that are
acceptable within the 90\% confidence limits.
The power-law index $\Gamma$ increased with the luminosity, as
expected from the negative correlation in the HIDs (figure
\ref{fig:hid}).
The Gaussian centroid for the iron line remained at
$E_\mathrm{Fe}=6.4$ keV throughout the period.
This appears inconsistent with the NuSTAR and NICER results that the
narrow ($\sigma\lesssim 300$ eV) iron-line centroid shifted from 6.4
to 6.7 keV in the luminous 
regime over the Eddington limit
\citep{2019ApJ...873...19T,2019ApJ...885...18J},
but this discrepancy is 
because the GSC spectrum with the
resolution $\Delta E\sim 0.8$ keV (at 6 keV) was dominated by the
broad structure with a peak at $\sim 6.4$ keV.
The equivalent width is almost constant at $EW_\mathrm{Fe}\sim 100$ eV,
in agreement with the NICER result that the iron-line flux was approximately 
proportional to the luminosity \citep{2019ApJ...885...18J},
as well as with the behavior of the typical XBPs \citep{2013A&A...551A...1R}.
When the 10 keV feature is fitted with a BB model,
the BB temperature increased from $kT_\mathrm{BB}\sim 1$ to 1.4 keV,
but the BB radius did not change significantly 
from $R_\mathrm{BB}\sim 10$ km. 
When it is fitted with the CYAB absorption model,
the CYAB energy and its width  
remained at $E_\mathrm{a}\sim 10$ keV and $W_\mathrm{a}\sim 3$ keV, respectively,
but the depth increased from $D_\mathrm{a}\sim 0.1$ to $0.2$ with the luminosity.

\subsubsection{Pulse-phase-resolved spectra}

Pulse profiles obtained by NICER in 0.2--12 keV 
were little energy dependent
during the luminous ($\gtrsim 2\times 10^{38}$ erg s$^{-1}$) period,
but their pulsed fractions
increased toward higher energies
\citepalias{2018ApJ...863....9W}.  
As seen in section \ref{sec:anapulse} (figure \ref{fig:pfrct}), the
same trend was observed in the GSC 2-20 keV data.  This suggests that
the X-ray spectrum gets harder around the pulse peaks.

\begin{figure}
\centering
\includegraphics[width=8.2cm]{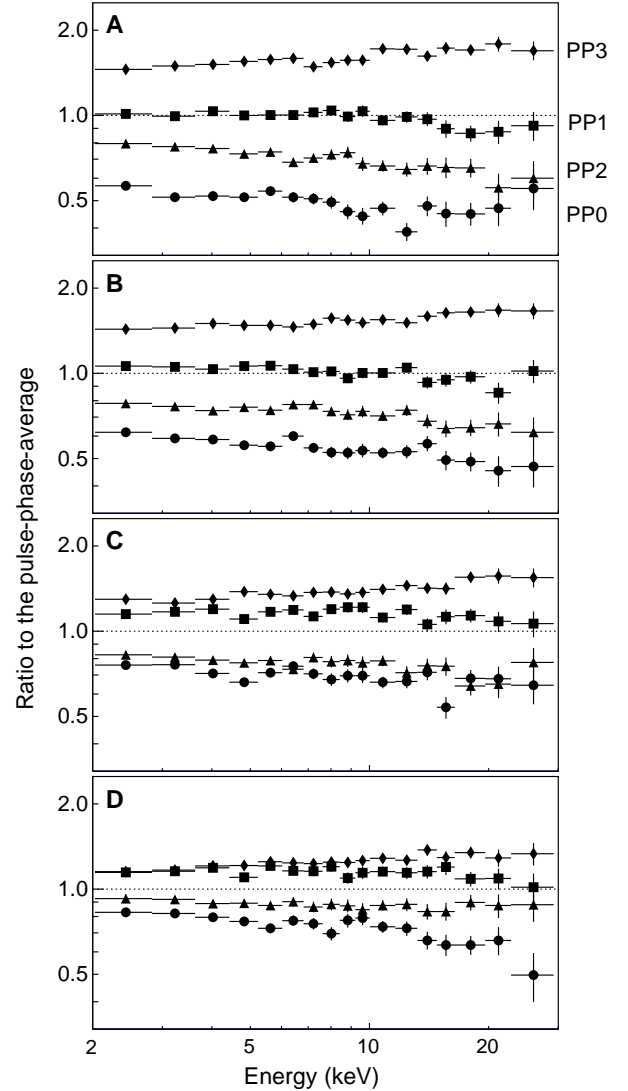} 
\caption{
Ratios of the pulse-phase-resolved spectra for PP0, PP1, PP2, and PP3 
to that of the pulse-phase average in the same interval.  
Penels A, B, C, and D, 
represents the spectra from the time intervals A to D, respectively.
\label{fig:ppspec_ratio}}
\end{figure}

We hence extracted pulse-phase-resolved spectra 
for 4 pulse phases (PP) as illustrated in figure \ref{fig:eflc},
which we hereafter call 
the minimum (PP1), 
the intermediate high (PP2), 
the intermediate low (PP3), 
and the maximum (PP4),
respectively, 
in the double-peak profile.
Figure \ref{fig:ppspec_ratio} shows
the ratios of each phase-resolved spectrum
to the entire phase average
during the luminous period of the intervals A, B, C, and D.
It confirms that the pulsed fraction indeed increases toward higher energies.
We also performed the model fit to the individual pulse-phase spectra,
but were not able to find significant phase-dependent parameter changes except for 
the power-law index $\Gamma$ and the emission normalization.

\subsection{Luminosity - spin-up relation}
\label{sec:ana_pdotlx}

As seen in figure \ref{fig:perhist},
the spin-frequency increase, i.e. the pulsar spin up,
is closely correlated with the X-ray intensity.
Although correlation was already reported by
\citet{2018A&A...613A..19D} and
\citet{2019ApJ...879...61Z},
we here refine the analysis by jointly
using the MAXI GSC light curve 
and the Fermi GBM pulse period.
These data have an advantage that 
both are available almost with
a daily sampling.

\begin{figure}
\centering
\includegraphics[width=8.2cm]{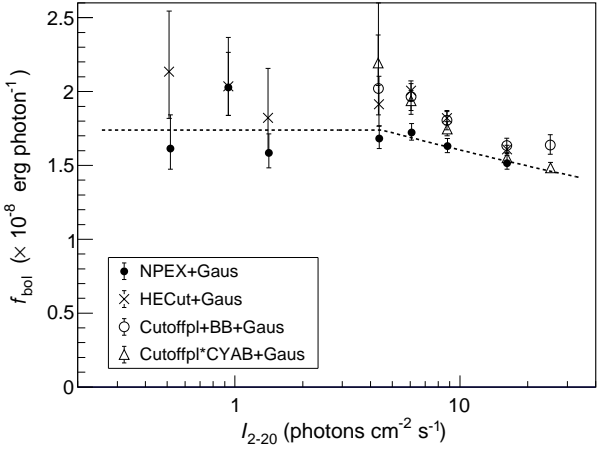} 
\caption{ 
Relation between $f_\mathrm{bol}=F_{0.5-60}/I_{2-20}$ and $I_{2-20}$,
calculated from the best-fit models  for the A through H spectra
in tables \ref{tab:specparam} and  \ref{tab:specparam2}.
Dashed line represents the best-fit function
to the NPEX-model data with equation (\ref{equ:fbol}).	  
\label{fig:convfact}}
\end{figure}

For the above purpose,
we need to convert $I_{2-20}$ to 
the bolometric luminosity $L_\mathrm{bol}$. 
The bolometric correction factor 
$f_\mathrm{bol} = F_{0.5-60}/I_{2-20}$, used in this conversion,
depends on the energy spectrum.
Figure \ref{fig:convfact} shows
the relation between $I_{2-20}$ and
$F_{0.5-60}$ calculated from the best-fit spectral models 
in tables \ref{tab:specparam} and  \ref{tab:specparam2}.
Although
the values of $F_{0.5-60}$ depend to some extent on the fitting models,
the effect is within the statistical uncertainties ($\lesssim 10\%$).
The factor $f_\mathrm{bol}$ 
slightly decreases towards the higher $I_{2-20}$,
according to the spectral softening
as observed in the HID (figure \ref{fig:hid}).
Based on the HID behavior, 
we assumed that the $f_\mathrm{bol}$-$I_{2-20}$ relatioin
can be expressed as 
\begin{equation} 
  f_\mathrm{bol} 
  = \left\{
  \begin{array}{ll}
    f_0                         & (I_{2-20}<4.5) \\
    f_0 \left(I_{2-20}/4.5\right)^{-\gamma}     & (I_{2-20}>4.5),
  \end{array} \right.
\label{equ:fbol}
\end{equation}
which is constant at $f_0$ in $I_{2-20}<4.5$ where HC is constant, 
and decreases by a power-law in $I_{2-20}>4.5$. 
We fitted equation (\ref{equ:fbol})
to the $f_\mathrm{bol}$-$I_{2-20}$ data obtained from the NPEX spectral parameters,
and determined the best-fit values of $f_0=1.74\times 10^{-8}$ erg
photon$^{-1}$, and $\gamma=0.10$.
The scale of $L_\mathrm{bol}$ in figures \ref{fig:hid}, \ref{fig:perhist}, \ref{fig:eflc} (a), 
and \ref{fig:pfrct} (b), associated with $I_\mathrm{2-20}$, 
have been calculated by 
$L_\mathrm{bol}=4\pi D^2 I_{2-20} f_\mathrm{bol}$ and $D=7$ kpc.

\begin{figure}
\centering
\includegraphics[width=8.2cm]{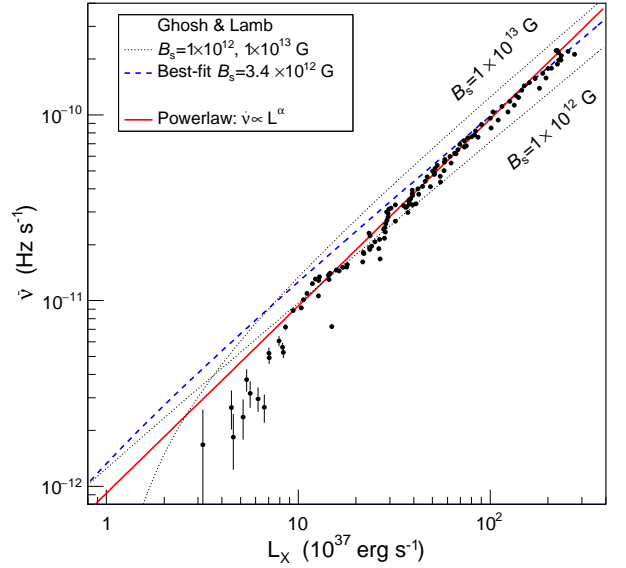} 
 \caption{The observed relation between
 $L_\mathrm{bol}$ and $\dot{\nu}_\mathrm{s}$.
 The dotted lines represent the \citetalias{Ghosh_Lamb1979} models 
 for surface magnetic field $B_\mathrm{s}=1\times 10^{12}$ and $1\times 10^{13}$ G.
 The red-solid and blue-dashed lines represent the best-fit power-law 
 and the best-fit \citetalias{Ghosh_Lamb1979} model, respectively.
 \label{fig:pdotlx}
 }
\end{figure}

Figure \ref{fig:pdotlx} shows the obtained 
$\dot{\nu}_\mathrm{s}$-$L_\mathrm{bol}$ relation,
where we calculated 
the spin-frequency derivative $\dot{\nu}_\mathrm{s}$ 
from the Fermi/GBM pulsar data
with the same procedure as in \citet{2017PASJ...69..100S}.
It clearly reveals a positive correlation close to the proportionality.
We fitted the data points with a power-law,
$\dot{\nu}_\mathrm{s} \propto  L_\mathrm{bol}^{\alpha}$,
and obtained the best-fit power-law index
$\alpha = 1.0~(\pm 0.02)$,
where the fitting error is estimated by 
adding appropriate systematic errors so as to make the fit formally acceptable.
The best-fit $\alpha$ value is somewhat higher than those of the theoretical predictions,
$6/7$ in 
\citeauthor{Ghosh_Lamb1979} (\citeyear{Ghosh_Lamb1979}, hereafter \citetalias{Ghosh_Lamb1979}),
$0.85$ in \citet{1995MNRAS.275..244L}, 
and $0.9$ in \citet{2007ApJ...671.1990K},
but agrees with the empirical relations determined 
from the observed data of
major Be XBPs \citep{1997ApJS..113..367B,2017PASJ...69..100S}.

We also compared the coefficient of proportionality between
$\dot{\nu}_\mathrm{s}$ and $L_\mathrm{bol}$, 
with those of the theoretical models.
Specifically, the date in figure \ref{fig:pdotlx}
are compared with the relations predicted 
by the representative \citetalias{Ghosh_Lamb1979} model,
assuming the canonical neutron-star mass $1.4 M_\odot$, 
the radius 10 km, 
the moment of inertial $10^{45}$ g cm$^{2}$, 
and the typical
surface-magnetic fields $B_\mathrm{s}=1\times 10^{12}$ and $1\times10^{13}$ G.
Although the data and the models slightly disagree in $\alpha$,
the data are mostly distributed between the two model curves.
This means that the data prefer $B_\mathrm{s}$ between these two values, 
i.e. a few $\times 10^{12}$.
The best-fit \citetalias{Ghosh_Lamb1979} model suggests $B_\mathrm{s}=3.4\times 10^{12}$ G.

\section{Discussion}

The MAXI GSC data of Swift J0243.6, during the giant outburst 
from 2017 October to 2018 January,
revealed the complex 
behavior in the X-ray spectrum as well as the pulse profile.
Based on these results, 
we consider possible scenarios of the X-ray emission evolution,
particularly at around the peak where the luminosity exceeded the Eddington limit
by up to a factor  of $\gtrsim 10$.
Also, comparing the behavior with those of other Be XBPs and ULXPs,
we discuss what causes the extraordinary super-Eddington emission of this object.

\subsection{Relations between spectral and pulse-profile transitions}
\label{sec:pulse_spec_relation}

The simultaneous changes in the X-ray spectrum and the pulse profile
of Swift J0243.6 have been noticed in the NICER and Fermi/GBM data
\citepalias{2018ApJ...863....9W}.
However,
possible relations between the two attributes have not
been necessarily clear, because of their uneven time coverage.
We here study this issue by using the MAXI GSC results.

As shown in figure \ref{fig:hid},
during the remarkable X-ray active period of $I_{2-20}\gtrsim 0.8$,
the two hardness rations, the SC and HC, 
both showed a negative correlation against $I_{2-20}$.
According to the simple HID classification 
\citep{2008A&A...489..725R},
it is classified into the diagonal branch (DB),
and the part  of $I_{2-20}\lesssim 0.8$
is thought to be the horizontal branch (HB)
from the result of NICER \citepalias{2018ApJ...863....9W}.
However, the two HIDs employing SC and HC 
show characteristic differences in the DB.
We hence divide the DB region into the follwoing two states,
(i) the intermediate DB state of $0.8\lesssim I_{2-20}\lesssim 4.5$
where the SC changed more than HC, 
and 
(ii) the extreme DB state of $I_{2-20}\gtrsim 4.5$
where the HC changed more than SC.
Using $f_\mathrm{bol}$ in equation \ref{equ:fbol},
these characteristic intensities of
$I_{2-20}=0.8$ and $4.5$ 
correspond to
the luminosities of $L_\mathrm{bol}=0.9\times 10^{38}$ and $5\times 10^{38}$ erg s$^{-1}$, 
respectively.

The spectral analysis clarified 
how the 2--30 keV spectrum changed between the two DB states.
Generally, X-ray spectra of Be XBPs are represented
with a Cutoffpl continuum \citep{Makishima1999, 2002ApJ...580..394C}, where
their luminosity-dependent changes in the DB are characterized by
a correlation between $L_\mathrm{bol}$ and $\Gamma$ 
\citep{2013A&A...551A...1R}.
As shown in figure \ref{fig:specparam_evolve},
the best-fit parameters obtained from Swift J0243.6
exhibit this general behavior. 
In the extreme DB state (intervals A, B, C, and D),
the increased 6 keV excess 
on top of the Cutoffpl continuum, 
further enhanced the change in the HC, but
reduced the change in the SC.

The pulse-profile evolution in figures \ref{fig:perhist} and
\ref{fig:eflc}
also suggests that it is related with the two DB states,
because transition between the single-peak and double-peak 
occurred at $I_{2-20}\simeq 4.5$,
just at the boundary of the two DB states.
These correlated changes in the spectrum and the pulse profile
are considered to reflect luminosity-related changes in
the physical condition of the X-ray emission region.
Table \ref{tab:transit} summarizes 
how the spectral and temporal properties
depend on the X-ray intensity.

\begin{deluxetable}{lcccccl}
\tablecaption{
Luminosity-dependent changes in the X-ray properties
\label{tab:transit}}
\tablehead{
{HID branch} & {HB} && \multicolumn{4}{c}{DB}\\
{Sub state in DB}  &                   && {Intermed.} && {Extreme}
}
\startdata
$I_{2-20}$$^{a}$            &&\multicolumn{2}{l}{0.8} & \multicolumn{2}{l}{4.5} & 30\\
$L_\mathrm{bol}$ ($10^{38}$)$^{b}$ &&\multicolumn{2}{l}{0.9} & \multicolumn{2}{l}{5.0} & 26 \\
Interval$^\mathrm{c}$           &   H     & \multicolumn{2}{c}{~~ G F (Y X)} & \multicolumn{2}{c}{~E D C B A} \\
\hline
SC-$I_{2-20}$ slope$^{d}$  &  $+$($\nearrow$) && $-$($\searrow$) && $\sim$$0$ ($\rightarrow$)\\ 
HC-$I_{-20}$ slope$^{d}$  &  $+$($\nearrow$) && $\sim$$0$ ($\rightarrow$)  && $-$($\searrow$)\\
Spec. profile           &  \multicolumn{4}{c}{Cutoffpl + Iron-K line} & \multicolumn{2}{l}{+ $\gtrsim6$ keV excess}\\
Pulse profile           &  \multicolumn{3}{c}{Single-peak} &&  \multicolumn{2}{l}{Double-peak} \\
\enddata
\tablecomments{
$^{a}$ 2--20 keV photon flux (photons cm$^{-2}$ s$^{-1}$).\\
$^{b}$ Bolometric luminosity (erg s$^{-1}$).\\
$^{c}$ GSC data intervals defined in table \ref{tab:outint}.\\
$^{d}$ $+$($\nearrow$) means positive correlation, and ($\searrow$) negative correlation, and $\sim$$0$($\rightarrow$) means little dependence.
}
\end{deluxetable}

\subsection{X-ray emission in the super-Eddington regime}
\label{sec:superedd_emission}

As discussed above,
the X-ray spectrum of Swift J0243.6 in the extreme DB state 
is characterized by the excess 
at $\gtrsim 6$  keV.
Because the feature can be represented by a Gaussian function 
with the centroid $\sim 6.4$ keV and 
the width $\sigma\sim 1.2$ keV,
\citet{2019ApJ...885...18J} interpreted it 
as a broad iron-K line.
However, a question about what cause such a broad iron line
has not been answered.
The broad Gaussian model also needs to have a large equivalent width of $\sim 1$ keV
\citep{2019ApJ...873...19T,2019ApJ...885...18J},
which would be realized only when the direct X-ray component 
is suppressed by source obscuration.
However, such an obscuration feature
has not been observed.
Meanwhile, 
to explain the power spectrum obtained from the insight-HXMT data
during the DB period,
\citet{2020MNRAS.491.1857D} proposed a scenario
that the major X-ray emission came from
an accretion disk, which made transition from
a state dominated by Coulomb collisions to that by radiation.
However, 
the picture is also considered difficult from
the pulsed fraction evolution,
which increased up to $\gtrsim 40$\% (RMS amplitude) 
in proportion to the luminosity. 
%
Furthermore, an accretion disk 
in a XBP must be truncated at 
the magnetospheric radius, or  
so-called Alfven radius \citep[][]{Ghosh_Lamb1979},
$R_\mathrm{A}=  1400 L_\mathrm{38}^{-2/7} M_{1.4}^{1/7} R_{6}^{10/7} B_{12}^{4/7} \, \mathrm{km}$,
where
$L_{38}$, $M_{1.4}$, $R_{6}$ and $B_{12}$ 
are the source luminosity in $10^{38}$ erg s$^{-1}$,
neutron-star mass in $1.4M_\odot$, 
radius in $10^6$ cm, 
and surface magnetic field in $10^{12}$ G. 
Therefore, the specific gravitational energy 
which the accreting matter acquires throughout the disk would be  
two order of magnitude smaller than is available by the time it reaches the neutron-star surface.
In other words, the disk would not provide a major source
for the observed pulsed hard X-rays.
The absorption line detected with the Chandra HETGS can be explained 
without invoking an X-ray emitting disk,
because the strong radiation pressure would produce
outflows from the cool disk outside $R_\mathrm{A}$,
or from the accretion stream inside $R_\mathrm{A}$.
%
Hence, we consider another interpretation
for these spectral and pulse-profile behavior.

As shown in figure \ref{fig:ratio2},
the MAXI GSC spectra with the excess feature
can be fitted if either 
a bump represented by a BB or an absorption by a CYAB model
is incorporated into the HECut or NPEX continuum.
These model parameters are consistent with
those for the "10 keV feature" which has been reported previously in several XBPs
\citep[e.g.][]{2002ApJ...580..394C, 2008A&A...491..833K}.
Also, similar spectral and pulse-profile changes
have been observed in several Be XBPs,
4U 0115$+$63 \citep{2009A&A...498..825F}, 
X 0331$+$53 \citep{2010MNRAS.401.1628T}, 
EXO 2030$+$375 \citep{2017MNRAS.472.3455E},
and SMC X-3 \citep{2017ApJ...843...69W},
when close to the Eddington limit.
These facts imply that the behavior is not unique to Swift J0243.6,
but common to the other XBPs.

Based on the canonical models of X-ray emission from XBPs 
\citep[e.g.][]{1976MNRAS.175..395B,2012A&A...544A.123B}, 
these X-rays are considered to originate from accretion columns
that are formed on the neutron star surface through the magnetic filed lines.
In this scenario, the two HID branches, HB and DB,
are thought to represent two accretion regimes 
where accreting matter flows 
are decelerated by Coulomb collisions (sub-critical accretion regime) and
radiation pressure (super-critical accretion regime),
respectively.
The spectral softening in the DB is interpreted by a development of 
Comptonized emission in the accretion columns.
As the luminosity increases, 
the region responsible for the Comptonization 
extends farther from the neutron star surface,
and then the temperature of the Comptonizing plasma 
decreases.
The scenario also explains the pulsed emission evolution
\citep{1975A&A....42..311B,2012A&A...544A.123B}.
Theoretically \citep{2012A&A...544A.123B}, 
the emission column height $h_\mathrm{s}$
is expected to be proportional to $L_\mathrm{bol}$ in the supercritical regime,
until it reaches a few km at the Eddington luminosity.
When $h_\mathrm{s}$  becomes larger than the column radius $r_\mathrm{c}$ ($\sim 1$ km),
the pulsed emission geometry changes
from pencil beam to fan beam,
which results in the transition
from the single-peak to the double-peak pulse profile.
Furthermore, if $h_\mathrm{s} \gg r_\mathrm{c}$,
the pulsed fraction tends to be approximately proportional to  $h_\mathrm{s}$,
and thus to $L_\mathrm{bol}$. 
The observed correlation between $f_\mathrm{pul}$ and $I_{2-20}$ in figure \ref{fig:pfrct} (b)
agrees with this prediction.

Then, what produces the 6 keV excess in the extreme DB state?
When the BB bump model is employed, 
the change of the feature with $L_\mathrm{bol}$ 
is represented by the BB temperature,
which increased from $kT_\mathrm{BB}=1.0$ to 1.6 keV.
On the other hand,
the BB radius was almost constant at $R_\mathrm{BB}\sim 10$ km.
Assuming that the BB emission came from the accretion column of $r_\mathrm{c}\sim 1$ km,
its height need to be $h_\mathrm{s}\sim 100$ km
to attain the BB area $=\pi R_\mathrm{BB}^2 \sim 100$ km$^2$.
The estimated $h_\mathrm{s}$ seems too high 
compared with the theoretical prediction of a few km.
This difficulty would not be solved even if we consider 
significant temperature gradient in the emission region.

Alternatively,
assuming the CYAB interpretation,
we obtained the best-fit parameters as
$E_\mathrm{a}\simeq 10$ keV, $W_\mathrm{a}\simeq 3$ keV,
and  $D_\mathrm{a}\simeq 0.1-0.2$.
Compared with other XBPs \citep[e.g.][]{Makishima1999, 2002ApJ...580..394C},
the values of $E_\mathrm{a}$ and $D_\mathrm{a}$ are at the lower ends of their distributions,
but still within their observed ranges.
The value of $W_\mathrm{a}$ is typical.
Therefore, the CYAB parameters are not so unusual.
In this scenario,
$E_\mathrm{a}= 10$ keV means 
$B_\mathrm{s}=0.86 (1+z_\mathrm{g})\times 10^{12}\simeq 1.1 \times 10^{12}$ G,
where $z_\mathrm{g}$ represents the gravitational redshift.
This estimate is consistent with the implication of figure \ref{fig:pdotlx}.
On the other hand,
the $L_\mathrm{bol}$ dependence of the parameters,
including an increase of $D_\mathrm{a}$ from 0.1 to 0.2,
and relatively constant values of $E_\mathrm{a}$ and $W_\mathrm{a}$,
are not necessarily typical of the cyclotron resonance effects 
in other XBPs, where $D_\mathrm{a}$ is relatively 
constant and $E_\mathrm{a}$ often decrease 
towards high $L_\mathrm{bol}$ \citep[e.g.][]{2004ApJ...610..390M}.
Therefore, we retain this interpretation as a possible candidate.

\subsection{Surface magnetic field}

The surface magnetic field $B_\mathrm{s}$ 
is one of the key parameters of the accretion process.
In the section above,
we arrived at a possibility of $B_\mathrm{s}\simeq 1.1 \times 10^{12}$ G,
assuming 
that the 6 keV excess feature in the spectrum of the extreme DB state
is a result of a cyclotron-resonance absorption at $\sim 10$ keV.
Meanwhile,
several other attempts to constrain $B_\mathrm{s}$ have been performed, so far.
\citet{2018MNRAS.479L.134T} 
derived $B_\mathrm{s}< 1\times 10^{13}$ G
from the upper limit on the propeller luminosity.
An estimate of $0.1-2 \times 10^{13}$ G was derived by \citetalias{2018ApJ...863....9W} 
from the HID transition luminosity and the QPO frequency.
From the correlated X-ray flux and spin-up evolution observed by the insight-HXMT,
\citet{2019ApJ...879...61Z} estimated $B_\mathrm{s}\sim 1\times 10^{13}$ G.
While all these constraints are consistent,
they have large uncertainties
which stem from those in the theoretical relations employed to 
interpret the observed data.
As a result, these published reports enable us to neither
assess the reality of our cyclotron hypothesis, nor examine
whether $B_\mathrm{s}$ of Swift J0243.6 is different from
those of typical XBPs.

We also studied this subject using 
the $\dot{\nu}_\mathrm{s}$-$L_\mathrm{bol}$ relation
from the MAXI/GSC and Fermi/GBM data (section \ref{sec:ana_pdotlx}),
and found that
the positive correlation between the two quantities
smoothly extends up to the maximum luminosity, 
$L_\mathrm{bol}\gtrsim 2\times 10^{39}$ erg s$^{-1}$
(figure \ref{fig:pdotlx}).
Assuming the neutron-star mass $1.4M_\odot$, 
the radius of 10 km, 
and the \citetalias{Ghosh_Lamb1979} disk-magnetosphere interaction model, 
the data are best explained with $B_\mathrm{s} \simeq 3.4\times 10^{12}$ G.
Although the model largely reproduce the data, the fit is 
not as good as being acceptable.
The discrepancy is considered mainly on 
the assumed physical conditions in \citetalias{Ghosh_Lamb1979},
which is estimated to affect the coefficient of proportionality 
between $\dot{\nu}_\mathrm{s}$ and $L_\mathrm{bol}$
by a factor of $\sim 2$ \citep[e.g.][]{2009A&A...493..809B}.
In fact, \citet{2017PASJ...69..100S} 
confirmed that the \citetalias{Ghosh_Lamb1979} model reproduced 
the observed $\dot{\nu}_\mathrm{s}$-$L_\mathrm{bol}$ relations of 12 Be XBPs
with an accuracy of a factor $\lesssim 3$.

\begin{figure*}
\centering
\includegraphics[width=18.cm]{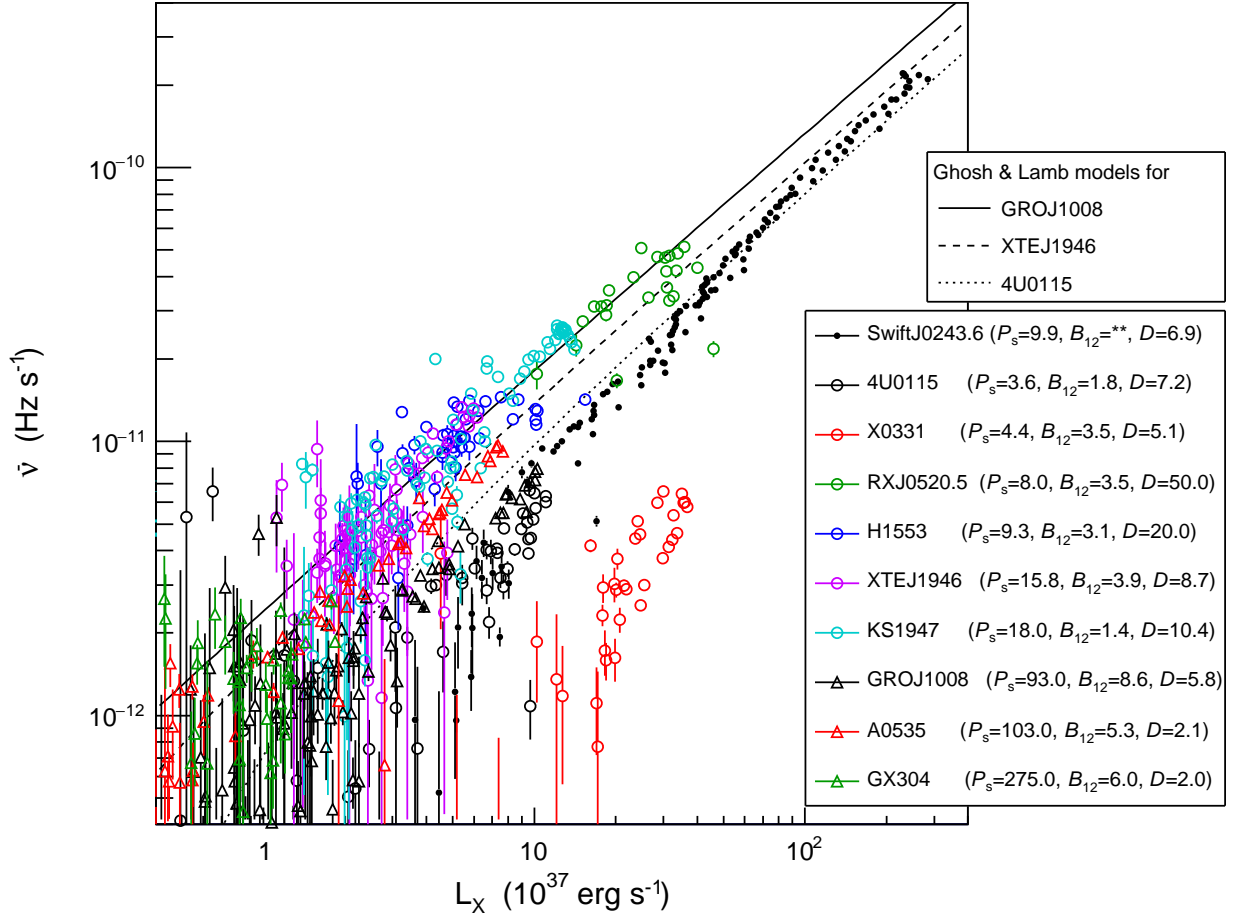} 
\caption{
   The observed $\dot{\nu}_\mathrm{s}$ versus $L_\mathrm{bol}$ relation of 
   Swift J0243.6 (black dot), compared with those of other 9 Be XBPs
   (4U 0115$+$63, X 0331$+$53, RX J0520.5$-$6932,
   H 1553$-$542, XTE J1946$+$274, KS 1947$+$300,
   GRO J1008$-$57, A 0535$+$262, and GX 304$-$1)
   whose $B_\mathrm{s}$ is determined by the cyclotron feature.
   All the data were produced from the MAXI GSC light curves and the Fermi GBM pulsar data \citep{2017PASJ...69..100S}.
   In the legend, the spin period $P_\mathrm{s}$ (s), 
   the surface magnetic field $B_{12}$ (in $10^{12}$ G), 
   and the assumed source distance $D$ (kpc) of each object
   are presented.
\label{fig:pdotlx_all}
}
\end{figure*}

To avoid these model uncertainties, 
we compare, in figure  \ref{fig:pdotlx_all},
the observed $\dot{\nu}_\mathrm{s}$-$L_\mathrm{bol}$ relation of Swift J0243.6
with those of other Be XBPs of which 
$B_\mathrm{s}$ is determined by the cyclotron-resonance feature.
These are the 9 Be XBPs in \citet{2017PASJ...69..100S};
4U 0115$+$63,
X 0331$+$53, 
RX J0520.5$-$6932,
H 1553$-$542,
XTE J1946$+$274,
KS 1947$+$300,
GRO J1008$-$57,
A 0535$+$262,
and GX 304$-$1.
The results for these XBPs have been derived 
from the MAXI/GSC and Fermi/GBM data in the same way as for Swift J0243.6.
The values of $L_\mathrm{bol}$ of 4 objects, 
4U 0115$+$63,
X 0331$+$53, 
A 0535$+$262,
and GX 304$-$1,
have been revised, 
using the updated $D$ in the GAIA DR2.
(These changes in $D$ from the values employed by \citet{2017PASJ...69..100S}
are $<15$\%.)
Except for one outlier, X 0331$+$53,
the $\dot{\nu}_\mathrm{s}$-$L_\mathrm{bol}$ relations 
of these objects all line up within a factor of $\sim 3$.
The data of Swift J0243.6
locate almost at the bottom of them,
in agreement with the fact that the best-fit \citetalias{Ghosh_Lamb1979} model
implies the lowest $B_\mathrm{s}$ among the known XBPs.
The result suggests that $B_\mathrm{s}$ of Swift J0243.6 is not
much different from the $B_\mathrm{s}$ range of XBPs,
and tends to be relatively low.
The timing analysis hence reinforce the cyclotron-absorption interpretation
of the $\sim 6$ keV excess feature.

\subsection{Comparison with other ULXPs}

In our Galaxy, 
Swift J0243.6 is the first example of the ULXP, 
as well as the ULX.
Therefore, the MAXI GSC results 
should give important hints about their unknown origins.
Table \ref{tab:ulxpsr} compares the basic  parameters of Swift J0243.6 
with those of the known 6 ULXPs,
M82 X-2 \citep{2014Natur.514..202B},
NGC 300 ULX-1 \citep{2018MNRAS.476L..45C},
NGC 7793 P13 \citep{2016ApJ...831L..14F},
NGC 5907 ULX-1 \citep{2017Sci...355..817I},
and SMC X-3 \citep{2017AA...605A..39T},
which have all been securely identified as ULXPs
with the maximum luminosities $>2.5\times 10^{39}$ erg s$^{-1}$.

\begin{deluxetable*}{llcclcccc}
\tablecaption{Basic parameters of known ULX pulsars
\label{tab:ulxpsr}}
\tablehead{
  \colhead{Source name}
  & \colhead{$P_\mathrm{spin}$ (s)}
  & \colhead{$P_\mathrm{orb}$ (d)}
  & \colhead{$L_\mathrm{max}$ (erg s$^{-1}$)}
  & \colhead{$D$ (Mpc)}
  & \colhead{Opt.}
  & \colhead{P/T}
  & \colhead{$B_\mathrm{s}^\mathrm{su}$ (G)}
  & \colhead{$B_\mathrm{s}^\mathrm{pr}$ (G)}
}
\startdata
NGC 5907 ULX-1$^{*1}$ & ~~1.14 & 5?    & $6.0\times 10^{40}$ & 17    & --  & P   & -- & -- \\
M82 X-2$^{*2}$        & ~~1.37 & 2.5   & $2.0\times 10^{40}$ & 3.5   & B9I & P   & -- & $\sim 1\times 10^{14}$\\
NGC 7793 P13$^{*3}$   & ~~0.42 & 64    & $1.0\times 10^{40}$ & 3.9   & --  & P   & $(1.5\times 10^{12}$) & --\\
NGC 300 ULX-1$^{*4}$  & 31.6   & --    & $5.0\times 10^{39}$ & 1.9   & Be  & T   & $3\times 10^{12}$ & --\\
SMC X-3$^{*5}$        & ~~7.8  & 45.1  & $2.5\times 10^{39}$ &   0.062     & Be  & T   & $2.6\times 10^{12}$ & $(1-5)\times 10^{12}$\\
Swift J0243.6$^{*6}$  & ~~9.7  & 27.6  & $2.5\times 10^{39}$ &  0.007    & Be  & T   & $2.5\times 10^{12}$ & $<6.2\times 10^{12}$ \\
\enddata
\tablecomments{
$P_\mathrm{spin}$ - spin period;
$P_\mathrm{orb}$ - orbital period;
$D$ - source distance;
$L_\mathrm{max}$ - observed maximum luminosity;
Opt. - optical counterpart;
P/T - Persistent or Transient;
$B_\mathrm{s}^\mathrm{su}$ - $B_\mathrm{s}$ from luminsity - spin-up relation;
$B_\mathrm{s}^\mathrm{pr}$ - $B_\mathrm{s}$ from propeller effect.
}
\tablerefs{
$^{*1}$\cite{2017Sci...355..817I},
$^{*2}$\cite{2014Natur.514..202B}; \cite{2016MNRAS.457.1101T},
$^{*3}$\cite{2016ApJ...831L..14F}, 
$^{*4}$\cite{2018MNRAS.476L..45C},
$^{*5}$\cite{2017AA...605A..39T},
$^{*6}$\cite{2018MNRAS.479L.134T}; \citetalias{2018ApJ...863....9W}.
}
\end{deluxetable*}

X-ray properties of XBPs depends considerably on 
the type of their mass-donating companions.
The XBPs known in our Galaxy
are mostly classified into
those accompanied by supergiant primaries, i.e. Sg XBPs, 
and the Be XBPs 
{\citep[e.g.][]{Reig2011,2015A&ARv..23....2W}}
which have been a major focus of the present paper.
While Sg XBPs show persistent X-ray activities
often involving flare-like time variations,
Be XBPs show mostly periodical outbursts lasting for a week to months
\citep[e.g.][]{1997ApJS..113..367B}.
Out of the 6 ULXP sample,
four have allowed optical identifications, and 
hence the classification; 
one Sg XBP and three Be XBPs.
From the type of their X-ray activity, the remaining two are naturally
considered to be Sg XBPs.
Therefore, regardless of its optical companion type,
any XBP may become, on certain conditions, an ULXP.
In table  \ref{tab:ulxpsr},
the ULXPs with Be companions are generally found to have
longer $P_\mathrm{s}$, as well as longer $P_\mathrm{orb}$,
than the objects of Sg companions,
in agreement with those of the XBPs in our Galaxy
\citep[][]{1986MNRAS.220.1047C}.
This suggests that the binary evolution of the ULXPs are not
much different from those of standard XBPs.

As the origin of the super-Eddington luminosity in ULXPs, 
a strong $B_\mathrm{s}$ reaching $\sim 10^{14}$ G has been
proposed with a theoretical model
\citep[][]{2015MNRAS.454.2539M}.  
However, it would be natured to presume that a stronger 
dipole filed would enlarge the Alfven radius and make it 
closer to the Bondi radius for gravitational capture
of the accreting gas, thus suppressing the accretion.
Actually, \citet{2018PASJ...70...89Y} 
found, through the $\dot{\nu}_\mathrm{s}$-$L_\mathrm{bol}$ technique,
that the very low-$L_\mathrm{bol}$ XBP, X Persei, 
has $B_\mathrm{s}\sim 10^{14}$ G.
Then, how about the values of $B_\mathrm{s}$ of the 6 ULXPs ?
In any of them, 
$B_\mathrm{s}$ has not been determined 
by the cyclotron feature.
Instead, its likely range has been estimated 
empirically and indirectly, employing
either; 
(a) the simultaneous luminosity - spin-up evolution \citep[e.g. this work; ][]{2018A&A...613A..19D,2019ApJ...879...61Z};
(b) the propeller effect \citep[e.g.][]{2016MNRAS.457.1101T,2017AA...605A..39T,2018MNRAS.479L.134T};
(c) assuming a torque equilibrium between the accreting matter and the pulsar magnetosphere \citep{2018MNRAS.476L..45C};
(d) the HID and/or pulse-profile transitions  \citep[][\citetalias{2018ApJ...863....9W}]{2017AA...605A..39T};
or (e) the QPO frequency \citepalias{2018ApJ...863....9W}.
Table \ref{tab:ulxpsr} refers to the results
obtained by (a) or (b),
because they are based on relatively simple theoretical models 
and have been better calibrated against observation data. 
Among these values, 
$B_\mathrm{s}^\mathrm{pr}\sim 10^{14}$ G in M82 X-2,
which was derived by \citet{2016MNRAS.457.1101T},
looks extraordinarily higher than the others ($\sim 10^{12}$ G).
However, 
M82 X-2 is consider to be a Sg XBP from the persistent X-ray activity,
so that its rapid flaring episodes could mimic the propeller effect.
All the other estimates agree with those of the standard XBPs,
$(1-8)\times 10^{12}$ G \citep[e.g.][]{Makishima1999,2014PASJ...66...59Y}.
This suggests that $B_\mathrm{s}$ of the ULXPs
are not different from those of the standard XBPs.

As discusse in sections \ref{sec:pulse_spec_relation} and \ref{sec:superedd_emission},
the X-ray behavior of Swift J0243.6 during the extreme DB state 
is represented by the spectral softening due to the broad 6-keV enhancement
and the transition from the single-peak to the double-peak pulse profiles.
Similar spectral and pulse-profile changes
at the luminosity close to the Eddington limit 
have already been reported in several Be XBPs,
even if they are not identified as UXLPs
\citep{2009A&A...498..825F,2010MNRAS.401.1628T,2017MNRAS.472.3455E,2017ApJ...843...69W}.
On the other hand, an absorption-like profile that can be fitted with a cyclotron-resonance
model,
was observed from another ULXP, NGC 300 ULX-1 \citep{2018ApJ...857L...3W}.
These results suggest that 
the observed properties in Swift J0243.6 during the extreme DB state
are common to ULXPs,
and smoothly extrapolated from those of the normal XBPs.

In summary, 
we find neither clear difference
between normal XBPs and ULXPs
in the basic parameters listed in table \ref{tab:ulxpsr},
and nor discontinuity in their luminosity-dependent X-ray behavior.
Therefore, the question, what causes the extraordinary high luminosity in ULXPs,
still remains unknown.
The key parameter might be in those that have not been discussed above.
One possible candidate would be
the angle $\theta_\mathrm{m}$ of the magnetic dipole moment
to the neutron-star spin axis. 
If $\theta_\mathrm{m}$  gets close to $90^\circ$,
the accretion path from the inner edge of the disk onto the neutron-star surface
through the field lines becomes shorter and more straight.
In the $\theta_\mathrm{m}\simeq 90^\circ$ case,
radiation pressure in the fan-beam geometery, which is expected 
under the super-critical accretion (section \ref{sec:superedd_emission}),
gets maximum in the direction perpendicular to
the accretion plane, and thus
it does not work effectively to decelerate the matter flow.
This mechanism will increase the maximum luminosity.

\section{Conclusion}

We analyzed the MAXI GSC data of the first ULXP in our Galaxy, 
Swift J0243.6, with a Be companion,
during the giant outburst from 2017 October to 2018 January.
The observed spectral and pulse-profile evolutions
during the extreme super-Eddington period
are explained by the scenario
that the accretion column responsible for the Comptonized X-ray emission
became taller as the luminosity increased.
One possible interpretation of the 6 keV excess feature,
which appeared significantly during the super-Eddington period,
is the presence of a cyclotron absorption feature at $\sim 10$ keV,
corresponding to $B_\mathrm{s}\simeq 1.1\times 10^{12}$ G.
The obtained $\dot{\nu}$-$L_\mathrm{L}$ relation close to the proportionality
is consistent with those of the standard Be XBPs
with $B_\mathrm{s}=(1-8)\times 10^{12}$ G.
The result thus suggests that $B_\mathrm{s}$ of Swift J0243.6 is a few $10^{12}$ G,
which is consistent with that implied by the cyclotron-absorption scenario.
Comparing the measured parameters
and the observed luminosity-dependent behavior
of the known 6 ULXPs including Swift J0243.6
with those of the standard XBPs,
we found no noticable difference.
Therefore, the key parameter to enable the super-Eddington accretion in XBPs
is yet to be identified.
The angle from the magnetic dipole moment to 
the neutron-star spin axis
would be one candidate.


\acknowledgments
The authors thank all the MAXI team members for their dedicated work on the mission operation.
Their thanks are also due to the Fermi/GBM
pulsar project for providing the useful results to the public.
This research has made use of 
data from the European Space Agency (ESA)
mission {\it Gaia} (\url{https://www.cosmos.esa.int/gaia}), processed
by the {\it Gaia} Data Processing and Analysis Consortium (DPAC,
\url{https://www.cosmos.esa.int/web/gaia/dpac/consortium}).
This work is partially supported by 
the Ministry of Education, Culture, Sports, Science and Technology (MEXT) of Japan
under Grants-in-Aid for
Science Research 17H06362 (M.S., N.K., and T.M.).
M.S. acknowledges support from the Strategic Pioneer Program on Space Science,
Chinese Academy of Sciences (grant No. XDA15052100).


\vspace{5mm}

\facilities{MAXI(GSC), Fermi(GBM)}
\software{
HEAsoft \citep[v6.25;][]{2014ascl.soft08004N},
XSPEC \citep[v12.8;][]{1996ASPC..101...17A} 
}




%
%
%

\end{document}